\documentclass{emulateapj}
\usepackage{graphicx}
\usepackage{amssymb}
\usepackage{lscape}
\usepackage[colorlinks = true, linkcolor = blue, urlcolor  = blue, citecolor = blue, anchorcolor = blue]{hyperref}
\usepackage{amsmath}
\usepackage{rotating}
\usepackage{natbib} 
\defcitealias{papovich16}{P16}
\begin{document}

\title{The \emph{Spitzer}-HETDEX Exploratory Large Area Survey II: Dark
  Energy Camera and \emph{Spitzer}/IRAC Multiwavelength Catalog}

\author{Isak G. B. Wold\altaffilmark{1,2},  Lalitwadee Kawinwanichakij\altaffilmark{3,4,5},
  Matthew L. Stevans\altaffilmark{1}, Steven
  L. Finkelstein\altaffilmark{1}, Casey Papovich\altaffilmark{3,4},
  Yaswant Devarakonda\altaffilmark{1}, Robin
  Ciardullo\altaffilmark{6,7}, John Feldmeier\altaffilmark{8},  Jonathan Florez\altaffilmark{1},  Eric
  Gawiser\altaffilmark{9}, Caryl
  Gronwall\altaffilmark{6,7}, Shardha Jogee\altaffilmark{1}, Jennifer L.
  Marshall\altaffilmark{3}, Sydney
  Sherman\altaffilmark{1}, Heath V. Shipley\altaffilmark{10}, Rachel
  S. Somerville\altaffilmark{9,11}, Francisco Valdes\altaffilmark{12}, and
Gregory R. Zeimann\altaffilmark{13}}

\altaffiltext{1}{Department of Astronomy, The University of Texas at
  Austin, 2515 Speedway, Stop C1400, Austin, Texas 78712, USA}
\altaffiltext{2}{NASA Goddard Space Flight Center, Greenbelt, MD, 20771, USA; isak.g.wold@nasa.gov}
 \altaffiltext{3}{Department of Physics and Astronomy, Texas A\&M University, College Station, TX, 77843, USA}
\altaffiltext{4}{George P. and Cynthia Woods Mitchell Institute for Fundamental Physics and Astronomy, Texas A\&M University, College Station, TX, 77843, USA}
\altaffiltext{5}{LSSTC Data Science Fellow}
\altaffiltext{6}{Department of Astronomy \& Astrophysics, The Pennsylvania State University, University Park, PA 16802, USA}
\altaffiltext{7}{Institute for Gravitation and the Cosmos, The Pennsylvania State University, University Park, PA 16802, USA}
\altaffiltext{8}{Department of Physics and Astronomy, Youngstown State University, Youngstown, OH 44555, USA}
\altaffiltext{9}{Department of Physics and Astronomy, Rutgers, The State University of New Jersey, 136 Frelinghuysen Rd, Piscataway, NJ 08854, USA}
\altaffiltext{10}{Department of Physics \& Astronomy, Tufts University, 574 Boston Avenue Suites 304, Medford, MA 02155, USA}
\altaffiltext{11}{Center for Computational Astrophysics, Flatiron Institute, 162 5th Ave, New York, NY 10010, USA}
\altaffiltext{12}{National Optical Astronomy Observatories, P.O. Box 26732, Tucson, AZ 85719, USA}
\altaffiltext{13}{Hobby Eberly Telescope, University of Texas, Austin, TX, 78712, USA}

\begin{abstract}
We present the $ugriz$-band Dark Energy Camera (DECam) plus 3.6 and 4.5 $\mu$m IRAC catalogs for
the \emph{Spitzer}/HETDEX Exploratory Large-Area (SHELA) survey. SHELA
covers $\sim24$ deg$^{2}$ of the Sloan Digital Sky Survey (SDSS)
``Stripe 82'' region, with seven bandpasses spanning a wavelength
range of 0.35 to 4.5 $\mu$m. SHELA falls within the
footprint of the Hobby-Eberly Telescope Dark Energy Experiment (HETDEX),
which will provide spectroscopic redshifts for $\sim200{,}000$ Ly$\alpha$
emitters at $1.9<z<3.5$ and also for $\sim200{,}000$ {[}O{\small{}II}{]}
emitters at $z<0.5$. SHELA's deep, wide-area multiwavelength images
combined with HETDEX's spectroscopic information, will facilitate many
extragalactic studies, including measuring the evolution of
galaxy stellar mass, halo mass, and environment from $1.5<z<3.5$.
 Here we present $riz$-band selected $ugriz$-band DECam catalogs that reach a $5\sigma$
depth of $\sim24.5$ AB mag (for point sources with an aperture that
encloses $70\%$ of the total flux) and cover $17.5$ deg$^{2}$ of the overall
SHELA field. We validate our DECam catalog by comparison to the DECam
Legacy Survey (DECaLS) DR5 and the Dark Energy Survey (DES) DR1. We
perform IRAC forced photometry with \textit{The Tractor} image
modeling code to measure 3.6 and 4.5 $\mu$m  fluxes for all objects
within our DECam catalog.  We demonstrate the utility of our catalog
by computing galaxy number counts and estimating photometric
redshifts. Our photometric redshifts recover the available
$\left\langle  z \right\rangle
= 0.33 $ SDSS
spectroscopic redshifts with a  $1\sigma$ scatter in $\Delta z/(1 +z)$ of 0.04.
\end{abstract}


\section{Introduction}

SHELA is a wide-field ($\sim24$ deg$^{2}$) multiwavelength survey
designed to study the intrinsic and environmental processes that govern the evolution of stellar mass in galaxies from
$1.5<z<3.5$. The survey's large area is needed to provide
statistically meaningful samples and to mitigate cosmic
variance. In the following, we explain SHELA's multi-wavelength
strategy with an emphasis on the role played by the Dark Energy Camera \citep[DECam,][]{honscheid08,flaugher15} imaging. The genesis and motivation of the SHELA survey has been thoroughly
discussed in our paper presenting the SHELA \emph{Spitzer} imaging
\citep[][P16 hereafter]{papovich16}. 

SHELA's  deep \emph{Spitzer}/IRAC \citepalias[][]{papovich16} and Mayall 4m/NEWFIRM
$K$-band images (Stevans et al.\ 2018, in prep) plus archival VICS82
survey $JK$-band images \citep{geach17} will provide accurate stellar
mass measurements for $M\gtrsim2\times10^{10}M_{\odot}$ galaxies
with known $1.5<z<3.5$ redshifts. While HETDEX \citep{hill08} will
provide spectroscopic redshifts for an important calibrating sample,
the vast majority ($\sim70\%$) of the IRAC selected galaxies from
$1.5<z<3.5$ will lack HETDEX detectable emission lines
  resulting in poorly constrained spectroscopic redshifts. Thus, deep
optical imaging combined with our redder bandpasses are needed to
measure photometric redshifts as well as probe the rest-frame ultraviolet.

To meet this requirement, we utilize DECam's wide field of view (3
deg$^{2}$) to obtain deep $ugriz$-band images covering $17.5$ deg$^{2}$
of the overall SHELA field. We note that our field is contained within
the DECam Legacy Survey (DECaLS\footnote{\url{http://legacysurvey.org/}})
which is a $grz$-band DECam survey covering 6700 deg$^{2}$ of the
equatorial sky. Although the nominal depth of DECaLS is $\sim1$ magnitude
shallower than our images, this collaboration plans to
incorporate all publicly available DECam datasets - including equatorial
Dark Energy Survey \citep[DES,][]{dark16, abbott18} DECam data and our SHELA DECam data. Thus, we
expect our $grz$-band DECam catalog depth to be comparable to DECaLS,
and we explore this topic in Section \ref{verifi}. We emphasize that DECaLS does not include
$i$-band data and -- critical to our photometric measurements -- does
not include $u$-band data. 

We need deep $u$-band imaging to accurately select $z>1.5$ galaxies
via their Lyman-break. Our SHELA $u$-band images are $1.5$ mag deeper
than the existing SDSS Stripe 82 $u$-band data which has a $5\sigma$
depth of $\sim23.9$ AB mag.  The importance of $u$-band imaging
to accurately measure photometric redshifts between $2\lesssim z\lesssim3.5$
is well known and has been discussed by many previous studies \citep[e.g.,][]{brunner97,brammer08}. As a proof of concept, in Section \ref{sphotoz} we compare our SHELA
$ugriz$$+JK+$IRAC photometric redshifts to the available $z<2$
SDSS spectroscopic data and show that we accurately recover spectroscopic
redshifts with a  $1\sigma$ scatter in $\Delta z/(1 +z)$ of 0.04. 

Rather than simply cross-matching our DECam catalog to the existing
SHELA IRAC catalog \citepalias[][]{papovich16}, we perform forced
photometry with \textit{The Tractor} image modeling code to measure
3.6 and 4.5 $\mu$m  fluxes for all objects within our DECam
catalog. This allows us to detect extremely faint sources that fall
well below the $5\sigma$ depth threshold of the original SHELA IRAC catalog
  \citepalias[22 AB mag][]{papovich16}.  Additionally, the improved
DECam resolution allows us to accurately measure fluxes for blended IRAC
sources.

In the following sections we discuss the DECam data reduction,
photometric depth estimates, construction of the catalog, computation
of number counts, and IRAC forced photometry. Unless otherwise noted, we give all magnitudes in
the AB magnitude system ($m_{\rm{AB}}=23.9-2.5 \log_{10}f_{\nu}$
with $f_{\nu}$ in units of $\mu$Jy).

\section{Observations}

We observed with the Blanco 4-m telescope at CTIO using the 3 deg$^{2}$
DECam imager and the $u$, $g$, $r$, $i$, and $z$-band filters.
DECam is a prime focus optical imager that has a focal plane array
consisting of sixty-two 2K$\times$4K pixel CCDs. Data were obtained
over 5 nights in the 2013B semester (PI: Papovich; NOAO PID: 2013B-0438), with one of these
nights lost due to weather. As shown in Figure \ref{tiles}, we observed
six slightly overlapping pointings covering a total area of $17.5$
deg$^{2}$. 

We supplement our data with overlapping DES data. DES DR1 is a DECam survey
that images $\sim5000$ square degrees in the $grizY$-bands to a 
10$\sigma$ depths of {[}$grizY${]}
$\sim$ {[}24.3, 24.1, 23.4, 22.7, 21.4{]} AB mag \citep{abbott18}.
The majority of DES lies at $\delta<-25$, but it has an extension
to cover Stripe 82 and our SHELA field. For the catalog presented
in this paper, we incorporate all the non-proprietary DES $griz$-band
images observed prior to October 2014 within the footprint of our
SHELA field. These DES images are also included in Figure \ref{tiles}
but are restricted to the footprint of our DECam survey. We also use Figure
\ref{tiles} to show the total exposure time in each tile and bandpass
(also see Section \ref{errsect} for  the area and median exposure time
of each tile).

\section{Data Reduction}

\begin{figure*}[!th]

\includegraphics[width=17.7cm]{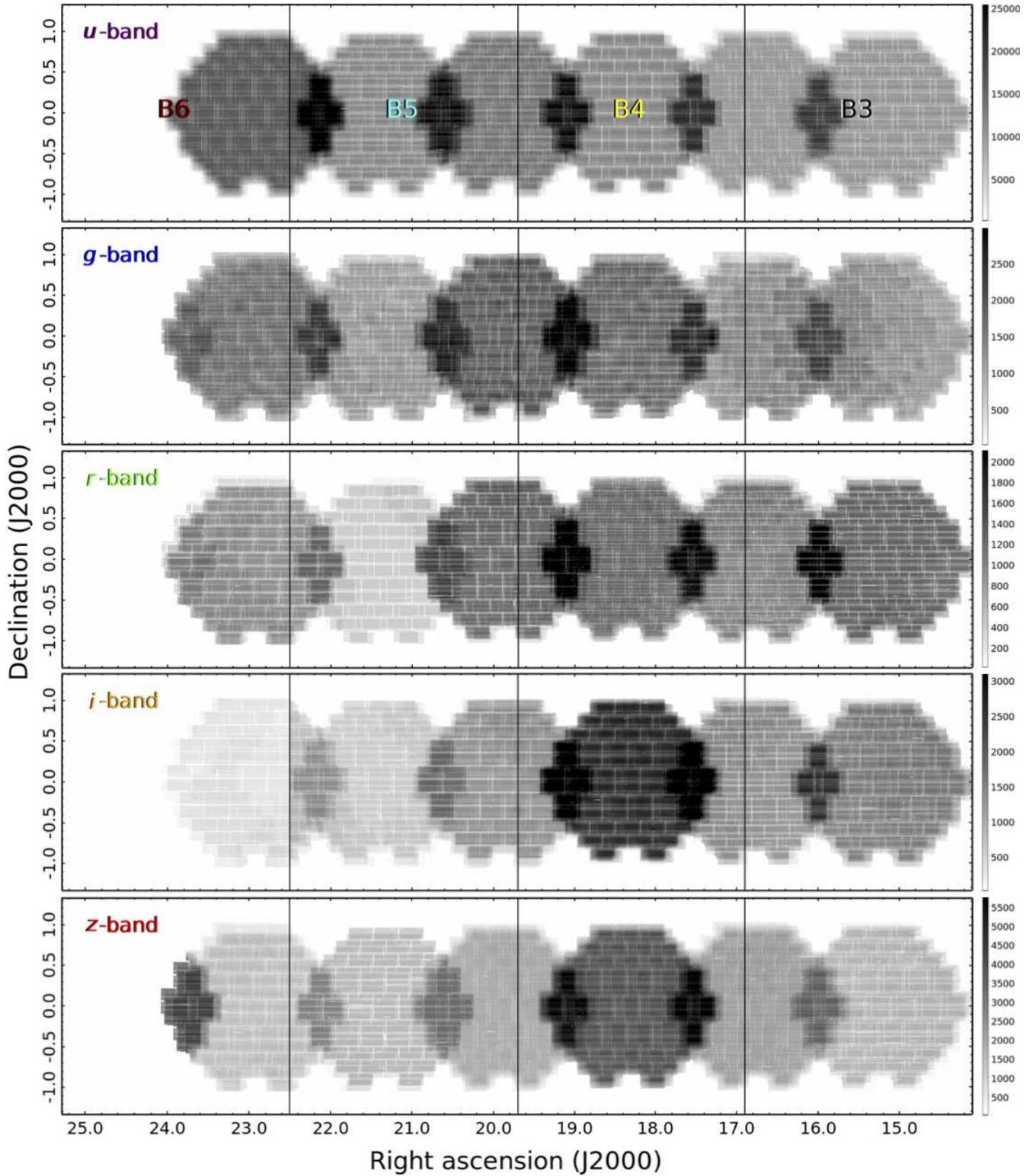}

\caption{SHELA $ugriz$-band exposure maps demonstrating the survey's relative
depth and area. The maps are divided into our four 2.8125 by 2.68125
degree tiles (B3, B4, B5, and B6; see top $u$-band panel). Right ascension and declination
are shown in degrees. The shaded vertical bars display the exposure
time in seconds. Our survey combines the $ugriz$-band DECam SHELA images with the overlapping $griz$-band DES images observed prior to October 2014.}

\label{tiles}

\pagebreak{}
\newpage\vspace{20mm}
\end{figure*}

We obtained all the non-proprietary NOAO Community Pipeline (CP; see
the NOAO Data Handbook\footnote{\label{note1}\url{http://ast.noao.edu/data/docs}})
reduced DECam $ugriz$ mosaics and associated data quality
maps (DQMs) observed prior to October 2014 within the SHELA survey
area from the NOAO Science Archive.  This consisted of data from both
our SHELA DECam survey and the DES.  For each band, we stacked the
available data into four 2.8125 by 2.68125 degree tiles, designated
B3, B4, B5, and B6. By design, each tile overlaps with its neighboring
tile by approximately 1 arc-minute. In this paper, we present the
DECam data over the region that has imaging in all 5 $ugriz$-bands.
In Figure \ref{tiles}, we show the field of view of our DECam images
and our tiling strategy.

\subsection{Image Stacking}\label{sect_reduc}
Our DECam data stacking procedure began with the NOAO DECam CP resampled
images. The NOAO DECam CP resampled images have been photometrically
and astrometrically calibrated with pixels aligned to a common 0.27
arcsec/pixel grid (see see the NOAO Data Handbook\footnotemark[\ref{note1}]
for details). The NOAO CP is an automated image reduction pipeline incorporating algorithms and elements from DES, IRAF, and earlier NOAO pipelines. The accuracy of the astrometry is limited by the science
exposure depth and the accuracy and overlap with the calibrating catalog,
which is currently the 2MASS catalog. The NOAO CP photometric calibration
uses the USNO-B1 catalog as the photometric reference and is not adequate
for science according to the NOAO Data Handbook. 

For these reasons, we recalibrate these resampled images to more closely
match the photometry and astrometry of SDSS. For each science image,
we first generate an initial SExtractor \citep[SE;][]{bertin96} catalog.
We use this catalog to determine the relative astrometry and to determine
the relative flux scaling factors required to adjust all images to
a common flux scale. We determine the relative astrometry by computing
the x- and y-offset required to match SDSS coordinates for `good'
stars. We define `good' stars as SDSS classified stars (class=6) with
a SDSS signal to noise ratio greater than 10 and no internal or external
SExtractor flags in our initial catalog. The typical offset we applied
was $\sim100$ mas (with a measured $1\sigma$ scatter of $\sim150$ mas),
which is within the quoted NOAO CP accuracy of $\sim200$ mas.

To determine the relative flux scaling factors, we must first measure
the total fluxes for isolated PSF stars. We define isolated PSF stars
as `good' stars with no neighbors within 5 arc-seconds in our
initial catalog. From
these isolated point sources, we determine the aperture that on median
encloses 70\% of the 5 arc-second aperture flux. We chose a  5
arc-second aperture diameter to mitigate flux contribution from
neighboring sources.  This 70\% aperture
flux is divided by a 0.7 correction factor to obtain the total flux
measurements. With total flux measurements for each exposure, we determine
the relative scaling factors required to scale all images to a designated
reference image. For each band, the reference image was selected from
tile B5 and at least 1,500 PSF stars were identified.  The flux scaling factors were typically centered
at unity with a standard deviation of $\sim0.1$.

We offset astrometry of each exposure to match SDSS and make an initial
photometrically scaled stack with SWarp \citep{bertin02}. We remove
any exposure with measured seeing worse than 2.5 arc-seconds from
the stack ($\sim2\%$ of the exposures meet this criteria). We also
remove a small number of images with obvious distortions or other
severe artifacts. This initial stack is weighted by:

\begin{equation}
w_{i}^{SB}=\left(\frac{1}{p_{i} \rm{rms}_{i}}\right)^{2}
\end{equation}

\noindent where $p_{i}$ is the flux scaling factor and rms$_{i}$ is the pixel-by-pixel
rms measured from background pixels in the science image. This weighting
scheme is the surface-brightness-optimized weighting discussed in
detail by \citet{gawiser06}. Any region assigned a non-zero DQM value
is given a weight of zero with one exception. We found that the NOAO
CP multi-exposure transient flagging procedure occasionally fails
and incorrectly flags the cores of bright stars. To mitigate this
problem we ignore all multi-exposure transient flags that overlap
with bright SDSS stars. We use the recommended Lanczos3 SWarp resampling
which preserves signal, while producing only modest artifacts around
image borders. 

\begin{figure}[!t]
\includegraphics[clip,width=8.5cm]{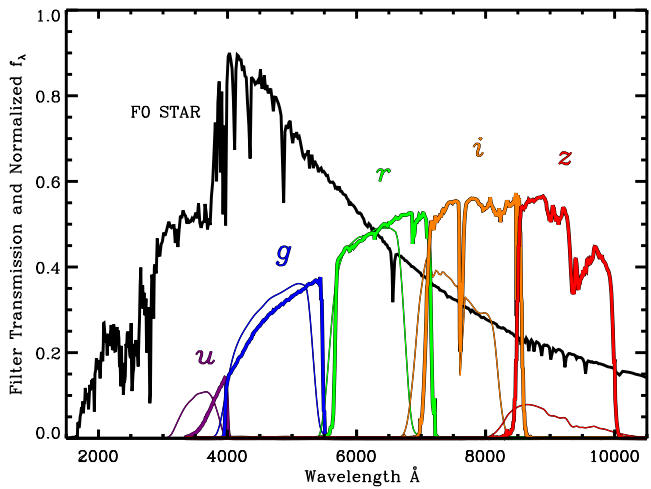}
\caption{Total system throughput for DECam filters (thick curves) and
  SDSS filters  (thin curves) and a normalized F0 star spectrum. }

\label{filters}

\end{figure}

With this initial stacked image, we determined the typical seeing,
FWHM$_{\rm{target}}$, and a seeing upper limit, FWHM$_{\rm{up}}$. The FWHM$_{\rm{target}}$
and FWHM$_{\rm{up}}$ are used to equalize the stacked point spread function.
Images with seeing less than FWHM$_{\rm{target}}$ are Gaussian smoothed
to achieve this target FWHM value. Images with seeing greater than
FWHM$_{\rm{up}}$ are excluded from the final stack. We note that our PSF
equalization procedure is very similar to the method used by SDFRED,
the standard Subaru Suprime-Cam reduction package \citep{ouchi04}.
The main advantage of this procedure is that it allows us to achieve
a uniform PSF without smoothing to the largest PSF size. Three members
of our team independently optimized the FWHM$_{\rm{target}}$ and the FWHM$_{\rm{up}}$
parameters to obtain a uniform PSF (isolated PSF stars typically have FWHMs that vary by
1$\sigma$ $\lesssim0\farcs05$). Our final reductions use the PSF parameters that achieved the best found image depth. In
practice the FWHM$_{\rm{up}}$ threshold was only needed and utilized on
8 of the 20 tile images.

The final stack is weighted by the point-source-optimized weighting
employed by \citet{gawiser06}:

\begin{equation}
w_{i}^{PS}=\left(\frac{s_{i}}{p_{i} \rm{rms}_{i}}\right)^{2}
\end{equation}

\noindent where $s_{i}$ is the seeing factor defined as:

\begin{equation}
s_{i}=1-exp\left(-1.3\frac{\rm{FWHM}_{\rm{target}}^{2}}{\rm{FWHM}_{i}^{2}}\right)
\end{equation}

\noindent  and FWHM$_{i}$ is the median FWHM measured for exposure $i$. This
stacking procedure results in 20 (4 tiles times 5 bandpasses) PSF-matched
science images and a corresponding number of weight maps (the sum
of input weights, $w_{i}^{PS}$). We emphasize that a target FWHM
must be known prior to producing a point-source-optimized stack, and
we estimate the target FWHM by first producing
a surface-brightness-optimized stack.  

\begin{deluxetable}{ccc} 
\tablecolumns{3} 
\tablewidth{0pc} 
\tablecaption{Photometric Calibration} 
\tablehead{ 
\colhead{Filter} & \colhead{F0 $\Delta m_{\rm{FILTER}}$} & \colhead{$\alpha$(SDSS AB color)}} 
\startdata 
$u$ & $0.329$ & \nodata \\
$g$ & $0.022$ & $0.122(g - r)$ \\
$r$ & $-0.002$ & $0.150(r - i)$ \\
$i$ & $-0.021$ & $0.303(i - z)$ \\
$z$ & $-0.009$ & $0.073(i - z)$
\enddata  
\tablecomments{F0 $\Delta m_{\rm{FILTER}}$ is the magnitude offset between SDSS and DECam filters for an F0 star. The $\alpha$ parameter is the color slope required to convert AB DECam magnitudes to SDSS magnitudes.}
\label{zpt1}
\end{deluxetable}

\subsection{Zero-point determination\label{szpt}}

We first derived the image zero-points using two different
determinations. In the first zero-point
determination method, we isolate F0 stars with a SDSS color cut. We
used F0 stars as a reference because of their large number density
and their relatively flat SED compared to lower-mass stars.  Using
SDSS filter transmission curves and a \citet{kurucz93} F0 spectral
energy distribution (SED), we determined the expected colors of a
F0 star. Using these colors, we constructed the following color cut
to achieve a large sample ($N>200$ per tile) of F0 stars: 

\begin{equation}
\begin{split}(u_{SDSS}-g_{SDSS}-0.96)^{2}+(g_{SDSS}-r_{SDSS}-0.14)^{2}\\
+(r_{SDSS}-i_{SDSS}+0.03)^{2}+(i_{SDSS}-z_{SDSS}+0.09)^{2}\\
<0.35^{2}
\end{split}
\end{equation}

The DECam filter set is not an exact match to the SDSS filter set
(see Figure \ref{filters}). For this reason, we computed the expected
magnitude offset between like filters ($\Delta m_{FILTER}$) assuming the available filter
transmission curves and a \citet{kurucz93} F0 SED. The zero-point
was then determined by:

\begin{equation}
ZPT_{F0}=\rm{median}(m_{SDSS}^{AB}-m_{DECam}-\Delta m_{FILTER})
\end{equation}

We report our $\Delta m_{FILTER}$ values in Table \ref{zpt1}. 

In the second zero-point determination
method, we examine all isolated PSF stars and assume a linear color
relation between SDSS and DECam magnitudes. For example, in the $g$-band
we assume:

\begin{equation}
g_{SDSS}^{AB}=g_{DECam}+ZPT_{Lin}+\alpha(g_{SDSS}-r_{SDSS})
\end{equation}

For tile B3, we solved for the zero-point and the color slope, $\alpha$,
by solving for the best-fit line. In subsequent tiles, we fixed $\alpha$
to the B3 value and solved for $ZPT_{Lin}$. We show below that this
fixed $\alpha$ assumption does not significantly affect our zero-point
determination. In Table \ref{zpt1}, we list the best-fit color slopes
and SDSS colors used for each band. In both methods the AB magnitude
for the DECam filter set is:

\begin{equation}
g_{DECam}^{AB}=g_{DECam}+ZPT_{F0,Lin}
\end{equation}

Our first method relies on the accuracy of the available filter profiles
and the model F0 SED, while our second method relies on there being
a linear color relation between SDSS and DECam magnitudes. For all
bands except the $u$-band (here many stellar SEDs display a sharp 4000
\AA\ break which makes the adjacent spectral slope a poor predictor of
the bluer DECam $u$-band flux) we find
a clear linear relation.    Thus,
we use the F0 zero-point determination for the $u$-band and use the
linear zero-point determination for all other bands. For $griz$-bands,
we find that the maximum $|ZPT_{F0}-ZPT_{Lin}|$ is 0.04 magnitudes
(see Table \ref{zpt2}). This provides an estimate for the systematic
uncertainty of our zero-points. We note that our zero-point accuracy,
as judged by $|ZPT_{F0}-ZPT_{Lin}|$, does not vary significantly
from one tile to the next, and this suggests that fixing $\alpha$
to the best-fit B3 value does not significantly affect our zero-point
determination.

We note that for AB magnitudes, sources with a flat flux density ($f_{\nu}$)
will have an AB color of zero. Thus, taking $g$-band as an example,
sources with an SDSS color term, $\alpha(g_{SDSS}-r_{SDSS})$, equal
to zero will have $g_{SDSS}^{AB}=g_{DECam}^{AB}$. We emphasize that
our cataloged magnitudes are AB for the observed DECam filters and
will in general differ from SDSS magnitudes by the SDSS color term,
$\alpha(g_{SDSS}-r_{SDSS})$. For all zero-point calculations, we
offset the SDSS $u$- and $z$-band magnitudes by $-0.04$ and $-0.02$,
respectively, to bring them in alignment with AB magnitudes\footnote{\url{http://classic.sdss.org/dr7/algorithms/fluxcal.html}}.

With zero-points accurately measured, we convert the image units to
nJy per pixel. This convention results in final image zero-points
of 31.4:

\begin{equation}
m_{\rm{AB}}=-2.5\log_{10}f_{{\rm {nJy}}}+31.4
\end{equation}

Our $grz$-band images are contained within DECaLS. DECaLS
will image 6700 deg$^{2}$ of the SDSS/BOSS extragalactic footprint
that lies in the region $-20<\delta<+30^{\circ}$ to depths of approximately
$g=24.7$, $r=23.9$, and $z=23.0$ AB mag (5-$\sigma$ point-source).
This survey intends to takes advantage of all non-DECaLS data that
lies within their survey area. Thus, the final DECaLS data release will incorporate data
from DES and our dataset and their $grz$-band
images should have comparable depth with our images (see Section \ref{verifi} for further discussion). Comparing their data
release 5 (DR5) cataloged $grz$-band data to our catalog of isolated PSF stars, we find
that our zero-points disagree by at most $0.05$ magnitudes (see Table
\ref{zpt2}). DECaLS's DR5 zero-points are computed from Pan-Starrs
PS1 photometry assuming a third order polynomial color relation between
PS1 and DECam magnitudes. We find that altering our calibration reference
catalog from SDSS to PS1 but keeping our linear color relation alters
our computed zero-points by $\sim0.02$ magnitudes. We attribute the
remaining SHELA / DECaLS zero-point offset to the different methods
employed to compute the zero-points (our linear adjacent color relation
vs.\ DECaLS' polynomial color relation). We decided to keep our linear
color relation method because of its use of adjacent color terms (which
should be a better measure of the local spectral slope), rather than
DECaLS's convention of using a $(g-i)$ color in all bands. 

\begin{deluxetable}{ccccc} 
\tablecolumns{5} 
\tablewidth{0pc} 
\tablecaption{Photometric Comparison} 
\tablehead{ 
\colhead{Tile} & \colhead{Filter} & \colhead{ZPT$_{\rm{F0}}-$ZPT$_{\rm{Lin}}$} & \colhead{ZPT$_{\rm{DECaLS}}-$ZPT$_{\rm{Lin}}$} & \colhead{ZPT$_{\rm{DES}}-$ZPT$_{\rm{Lin}}$}} 
\startdata 
B3 & $u$ & \nodata & \nodata & \nodata \\
  & $g$ & $0.03\pm0.08$ & $0.02\pm0.03$ & $0.04\pm0.03$ \\
  & $r$ & $0.01\pm0.07$ & $-0.05\pm0.05$ & $-0.00\pm0.04$ \\
  & $i$ & $0.04\pm0.08$ & \nodata & $0.03\pm0.03$ \\
  & $z$ & $0.01\pm0.09$ & $-0.05\pm0.03$ & $0.02\pm0.03$ \\\hline
\rule{0pt}{4ex}B4 & $u$ & \nodata & \nodata & \nodata \\
  & $g$ & $0.03\pm0.08$ & $0.01\pm0.03$ & $0.03\pm0.03$ \\
  & $r$ & $0.01\pm0.07$ & $-0.04\pm0.02$ & $-0.00\pm0.03$ \\
  & $i$ & $0.04\pm0.07$ & \nodata & $0.03\pm0.02$ \\
  & $z$ & $0.01\pm0.09$ & $-0.04\pm0.04$ & $0.02\pm0.04$ \\\hline
\rule{0pt}{4ex}B5 & $u$ & \nodata & \nodata & \nodata \\
  & $g$ & $0.03\pm0.08$ & $0.01\pm0.03$ & $0.03\pm0.03$ \\
  & $r$ & $0.01\pm0.06$ & $-0.04\pm0.02$ & $-0.01\pm0.03$ \\
  & $i$ & $0.04\pm0.09$ & \nodata & $0.04\pm0.04$ \\
  & $z$ & $0.01\pm0.09$ & $-0.05\pm0.03$ & $0.02\pm0.02$ \\\hline
\rule{0pt}{4ex}B6 & $u$ & \nodata & \nodata & \nodata \\
  & $g$ & $0.03\pm0.08$ & $0.01\pm0.02$ & $0.03\pm0.03$ \\
  & $r$ & $0.01\pm0.07$ & $-0.03\pm0.02$ & $0.00\pm0.02$ \\
  & $i$ & $0.03\pm0.07$ & \nodata & $0.03\pm0.02$ \\
  & $z$ & $0.00\pm0.10$ & $-0.05\pm0.02$ & $0.02\pm0.02$
\enddata  
\tablecomments{ZPT$_{\rm{F0}}-$ZPT$_{\rm{Lin}}$ is the median zero-point offset between our F0 and our linear color determination method. ZPT$_{\rm{DECaLS}}-$ZPT$_{\rm{Lin}}$ is the median zero-point offset between DECaLS and our linear color determination method. ZPT$_{\rm{DES}}-$ZPT$_{\rm{Lin}}$ is the median zero-point offset between DES and our linear color determination method. The errors indicate 1.48 times the median absolute deviation.}
\label{zpt2}
\end{deluxetable}

Recently, DES has published DR1 \citep{abbott18} with a reported photometric precision of $<1\%$ in all bands.  
Comparing their DR1 cataloged $griz$-band data to our catalog of isolated PSF stars, we find
that our zero-points disagree by at most $0.04$ magnitudes (see Table
\ref{zpt2}).  Both DES and DECaLS suggest  that our $g$-band zeropoint is systematically offset by $\sim3\%$.  However, DES agrees with our $r$ and $z$-band zeropoints ($\le2\%$), while DECaLS suggests a $\sim4\%$ correction.  Given these issues, we leave our zeropoints unchanged and adopt a $5\%$ systematic uncertainty in our cataloged flux errors (See Section \ref{errsect}).

\subsection{Final PSF Matching\label{sect_psf}}

\begin{figure}[!t]
\includegraphics[width=8.5cm]{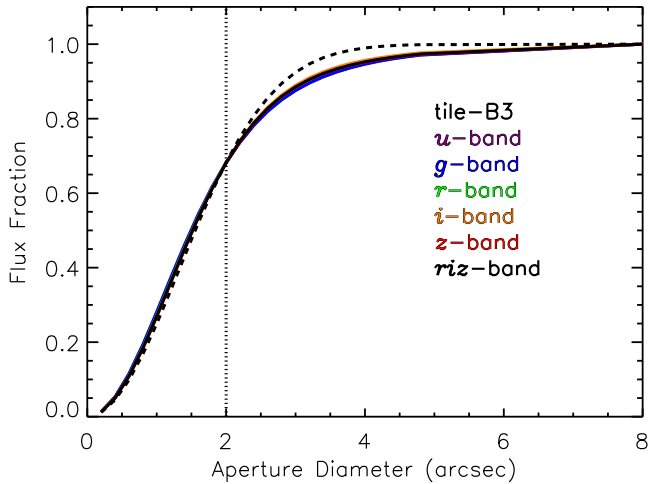}
\caption{Median curve of growth for isolated stars in the PSF-matched B3 tile.
We show curves of growth for our $ugriz$-band and for our detection
$riz$-band image. For comparison, we also show a Gaussian profile
with the dashed black curve. The dotted vertical line indicates the
$70\%$ flux aperture used to compute the total flux for point sources.
For our completeness simulations, we insert fake Gaussian sources
(profile shown by dashed black curve) with FWHM set to match the measured
$70\%$ curve of growth flux fraction. (The complete figure set (4
images) for all SHELA tiles is available in the online journal.) }

\label{cog}

\end{figure}

\begin{deluxetable*}{ccccccccc} 
\tablecolumns{9} 
\tablewidth{0pc} 
\tablecaption{Seeing, Apertures, and Point Source Detection Limits.} 
\tablehead{ 
\colhead{Tile} & \colhead{Filter} & \colhead{Input $2\times r^{SE}_{1/2}$} & \colhead{Output $2\times r^{SE}_{1/2}$ } & \colhead{PSF} & \colhead{Flux} & \colhead{Sky Aperture}  & \colhead{Simulation}& \colhead{DECaLS DR5}\\ 
\colhead{ } & \colhead{ } & \colhead{Median} & \colhead{Median} & \colhead{Aperture} & \colhead{Enclosed} & \colhead{Detection Limit}  & \colhead{Detection Limit} & \colhead{Detection Limit}\\ 
\colhead{ } & \colhead{ } & \colhead{(arcsec)} & \colhead{(arcsec)} & \colhead{(arcsec)} & \colhead{(Fractional)} & \colhead{($5\sigma$, AB)} & \colhead{($5\sigma$, AB)} & \colhead{($5\sigma$, AB)}}
\startdata 
B3 & $u$ & 1.38 & 1.48 & 2.0 & 0.68 & 25.4 & 25.1 & \nodata \\
  & $g$ & 1.47 & 1.47 & 2.0 & 0.68 & 25.1 & 24.9 & 25.2 \\
  & $r$ & 1.37 & 1.49 & 2.0 & 0.68 & 24.8 & 24.7 & 24.9 \\
  & $i$ & 1.18 & 1.50 & 2.0 & 0.68 & 24.3 & 24.1 & \nodata \\
  & $z$ & 1.14 & 1.49 & 2.0 & 0.68 & 23.9 & 23.6 & 23.8 \\
  & $riz$ & \nodata & 1.49 & 2.0 & 0.68 & 25.0 & 24.8 & \nodata \\\hline
\rule{0pt}{4ex}B4 & $u$ & 1.41 & 1.59 & 2.2 & 0.71 & 25.4 & 24.9 & \nodata \\
  & $g$ & 1.54 & 1.54 & 2.2 & 0.71 & 25.1 & 24.8 & 25.1 \\
  & $r$ & 1.34 & 1.58 & 2.2 & 0.71 & 24.7 & 24.5 & 24.9 \\
  & $i$ & 1.15 & 1.59 & 2.2 & 0.71 & 24.2 & 24.0 & \nodata \\
  & $z$ & 1.42 & 1.56 & 2.2 & 0.71 & 23.7 & 23.5 & 23.8 \\
  & $riz$ & \nodata & 1.57 & 2.2 & 0.71 & 24.7 & 24.6 & \nodata \\\hline
\rule{0pt}{4ex}B5 & $u$ & 1.43 & 1.56 & 2.2 & 0.71 & 25.4 & 25.0 & \nodata \\
  & $g$ & 1.54 & 1.54 & 2.2 & 0.71 & 25.0 & 24.8 & 25.2 \\
  & $r$ & 1.34 & 1.58 & 2.2 & 0.71 & 24.4 & 24.3 & 24.7 \\
  & $i$ & 1.26 & 1.58 & 2.2 & 0.71 & 23.8 & 23.5 & \nodata \\
  & $z$ & 1.21 & 1.59 & 2.2 & 0.71 & 23.7 & 23.5 & 23.5 \\
  & $riz$ & \nodata & 1.59 & 2.2 & 0.71 & 24.6 & 24.5 & \nodata \\\hline
\rule{0pt}{4ex}B6 & $u$ & 1.52 & 1.54 & 2.2 & 0.71 & 25.3 & 24.9 & \nodata \\
  & $g$ & 1.53 & 1.53 & 2.2 & 0.71 & 25.0 & 24.8 & 25.1 \\
  & $r$ & 1.37 & 1.55 & 2.2 & 0.71 & 24.5 & 24.4 & 24.8 \\
  & $i$ & 1.31 & 1.57 & 2.2 & 0.71 & 23.7 & 23.4 & \nodata \\
  & $z$ & 1.23 & 1.58 & 2.2 & 0.71 & 23.6 & 23.4 & 23.8 \\
  & $riz$ & \nodata & 1.56 & 2.2 & 0.71 & 24.6 & 24.5 & \nodata
\enddata  
\tablecomments{The half-light radius ($r^{SE}_{1/2}$) is the radius containing $50\%$ of the total light, where the total light is measured within SExtractor's AUTO aperture. Twice the half-light radius is the half-light diameter, which is a robust estimate of the seeing FWHM \citep[e.g.,][]{gawiser06}. As described in Section \ref{sect_psf}, the `PSF Aperture'  column shows the circular aperture diameter that contains approximately $70\%$ of the total light for points sources, where the total light is measured by a $8''$ aperture diameter. The `Flux Enclosed' column shows the precise amount of flux enclosed within our PSF aperture.  The tabulated sky aperture and simulation $5\sigma$ detection limits are descibed in Section \ref{errsect} and illustrated in Figure \ref{uncert}. The DECaLS DR5 detection limits are descibed in Section \ref{verifi}.}
\label{btable}
\end{deluxetable*}

\begin{deluxetable}{lc} 
\tablecolumns{2} 
\tablewidth{0pc} 
\tablecaption{SHELA SExtractor Parameter Settings} 
\tablehead{ 
\colhead{SExtractor Parameter} &  \colhead{Value}} 
\startdata 
DETECT\_MINAREA & 3 pixels \\
DETECT\_THRESH & 1.5 \\
ANALYSIS\_THRESH & 1.5 \\
FILTER\_NAME & Gauss\_3.0\_7x7.conv \\
WEIGHT\_TYPE & MAP\_RMS,MAP\_RMS \\
DEBLEND\_NTHRESH & 32 \\
DEBLEND\_MINCONT & $10^{-5}$ \\
MAG\_ZEROPOINT & $31.4$ \\
PIXEL\_SCALE & $0.27$ arcsec \\
BACK\_TYPE & AUTO,AUTO \\
BACK\_SIZE & $256$ \\
BACK\_FILTERSIZE & $4$  \\
MASK\_TYPE & CORRECT  \\
CLEAN & NO 
\enddata  
\label{se}
\end{deluxetable}

For each tile, we degrade the PSF to match the measured PSF in the
bandpass with the worst seeing. We perform this final Gaussian PSF
matching to ensure that an aperture with fixed size will enclose the
same fraction of a point-source's flux regardless of the observed
bandpass. In particular, we wish to scale up a fixed $70\%$ aperture
flux to determine the total flux of point sources. In Figure \ref{cog},
we show the median fraction of flux enclosed vs.\ aperture diameter
for isolated stars in the B3 tile. In Table \ref{btable}, we report
the half-light diameter for isolated stars before and after final
PSF matching for all bandpasses and tiles. In Table \ref{btable},
we also report the our adopted $70\%$ aperture diameter and the precise
amount of flux enclosed within this aperture. We have adopted the
convention of sampling aperture diameters every $0\farcs2$, thus the
closest PSF aperture will enclose slightly more or less flux than
$70\%$. The cataloged total fluxes are adjusted accordingly. For
example, we compute the total AB $u$-band magnitude in tile B3 via:

\begin{equation}
u_{AB}^{tot}=u_{AB}^{2\farcs0}+2.5\log_{10}0.68
\end{equation}
We emphasize that our total PSF fluxes are only accurate for point
sources. For resolved objects, we catalog SExtractor MAG\_AUTO magnitudes.
SExtractor's AUTO aperture uses a Kron ellipse adjusted to each object's
light profile and is meant to give the most precise SExtractor estimate
of total magnitude.

\begin{figure}[!t]
\includegraphics[width=8.5cm]{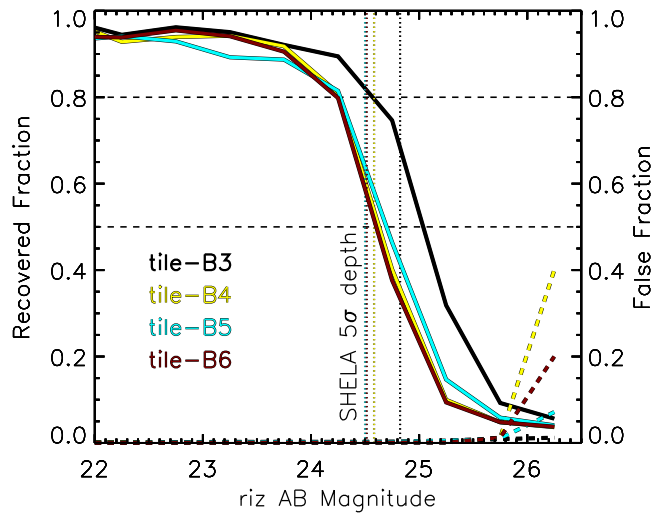}
\caption{Recovered fraction as a function of input $riz$-magnitude for simulated
point sources. We added simulated points sources with no color ($u=g=r=i=z$)
to all of the science images and then employed our normal source extraction
procedure. The solid red, orange, green, and blue curves indicate
the recovered fraction of simulated sources for tiles B3, B4, B5,
and B6, respectively. The dashed curves indicate the estimated false
detection fraction as a function of magnitude. We inverted the detection
image and measured the number of negative sources extracted per magnitude
bin. We took the ratio of the number of negative sources to the number
of sources extracted from our science images to estimate false detection
fraction. We find a $50\%$ ($80\%$) completeness limit at $\sim24.7$ ( $\sim24.4$) mag. False sources only become significant at
magnitudes fainter than our estimated $10\%$ completeness limit. We indicate our catalog's $5\sigma$ depth with
  vertical dotted lines, see Section \ref{errsect} for details.}

\label{complete}

\end{figure}

\subsection{Detection Image and Source Extraction\label{sect_se}}

For each tile, we compute an inverse variance-weighted summation of
the $r$, $i$, and $z$-bands to construct a detection image. The
variance image is formed by normalizing our inverse weight maps (described
in  Section  \ref{sect_reduc}) to the variance measured in our science
frames prior to final PSF matching. For each tile
all images were scaled to a common nJy scale and
have approximately the same PSF. Thus, our inverse variance weighting
is consistent with the point-source optimized weighting employed by
\citet{gawiser06}. We decided not to use the $u$ and $g$-bands in our
detection image because many high-redshift ($z>3.5$) galaxies will dropout of
these bands, so their inclusion would only add noise for these sources.

 In Figure \ref{cog}, we compare the detection
image's PSF (black \emph{solid} curve) to the PSF in other bandpasses
(purple, blue, green, orange, and red curves) using tile B3 as an
example. As expected the agreement is good, and in Table \ref{btable},
we quantify this agreement by showing that the same flux fraction
is enclosed within our adopted PSF aperture for both our detection
image and our $ugriz$-bandpass images. 

We used SExtractor in double-image mode to detect and measure sources
with the SExtractor parameters listed in Table \ref{se}. In this
mode the first image specified is used for the detection of sources
and the second image is used for measuring source properties. This
allowed us to produce uniform $riz$-selected $ugriz$-band catalogs.
We utilize DQMs that have non-zero pixels when all pre-stacked pixels
are flagged in the NOAO CP DQMs. We also assign non-zero DQM values
to regions around bright stars. Bright stars effectively mask out
regions of the sky and have associated false sources that are due
to diffraction spikes and saturation effects. We mitigate these issues
by masking out all UCAC4 sources \citep{zacharias13} with magnitudes
less than 16.5. We determined a magnitude dependent circular star
mask defined by:

\begin{equation}
{\rm {Radius}}=319.7-41.9R+1.4R^{2}\label{eq:10-1}
\end{equation}

\begin{figure}[]
\includegraphics[width=8.5cm]{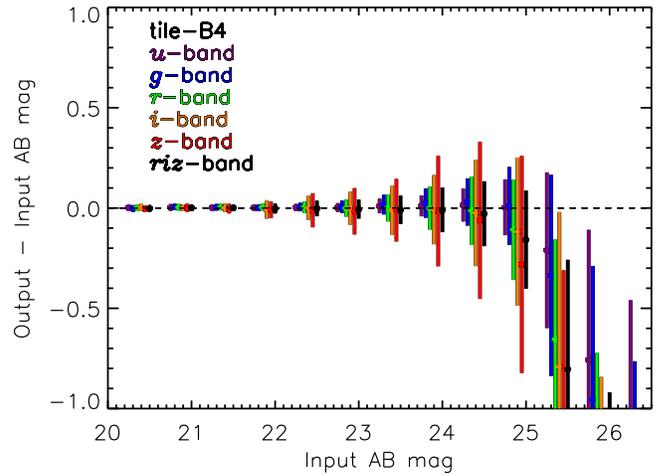}
\caption{Median difference between measured output and known input magnitude
of fake sources as a function of input magnitude for tile B4. The error bars indicate
1.48 times the median absolute deviation ($\sigma_{{\rm {MAD}}}$).
For clarity data points are slightly offset in the x-direction from
the unaltered position occupied by the u-band data points. The median difference between the input and
recovered magnitude is near zero for objects brighter than $\sim24.5$
mag.  We use these magnitude errors to estimate our catalog's
$5\sigma$ depth, see Section \ref{errsect} and Figure \ref{btable}. (The complete figure set (4
images) for all SHELA tiles is available in the online journal.) }

\label{inout}

\end{figure}

\noindent where the radius is in arcsec and $R$ is the UCAC4 fit model magnitude
with bandpass spanning $5790-6420$ \AA. We constructed this star
mask by adjusting circular regions centered on UCAC4 stars to include
all obvious false sources (e.g., sources associated with diffraction
spikes) for a sample of stars with a wide range of magnitudes. With
these radii and magnitudes, we fit a polynomial of degree 2, resulting
in Equation \ref{eq:10-1} \citep[for a similar procedure see][]{keenan10}.
We record both the internal (see SExtractor User's Manual) and DQM-based external SExtractor flags in our
final catalog.

\begin{figure}[!t]
\includegraphics[width=8.5cm]{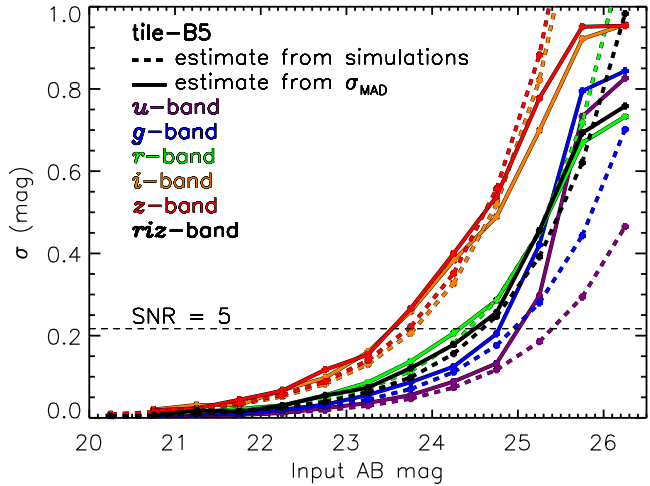}
\caption{Estimated photometric uncertainties determined from our simulations
(solid curves) and from the noise measured in $70\%$ apertures scaled
to total magnitudes (dashed curves) for tile B5. The black dashed line indicates
the magnitude uncertainty that corresponds to a SNR$=5$. The intersection
of the SNR$=5$ dashed line and the photometric uncertainty curves
gives our catalog's $5\sigma$ depth; recorded in Table \ref{btable}
for all tiles. (The complete figure set (4
images) for all SHELA tiles is available in the online journal.) }

\label{uncert}

\end{figure}

We also record SExtractor morphological information including the semimajor axis
($a$), the ellipticity ($e=1-b/a$, where $b$ is the semiminor axis),
the position angle of the semimajor axis in degrees east from celestial
north ($\theta$), the Kron radius ($r_{Kron}$), and the half-light
radius ($r_{1/2}$).


\subsection{Completeness and Purity tests \label{simsect}}

For each tile, we inserted 10,000 fake Gaussian sources with no color
($u=g=r=i=z$) into our final science images and then attempted to
recover them using our extraction procedure to estimate the completeness
of our catalog. We set the FWHM of the Gaussian sources to match the
enclosed flux within our PSF aperture. In Figure \ref{cog}, we show
 (with a black dashed line) the curve of growth for our fake Gaussian sources used in the B3 tile. For each fake source, we randomly assigned an image position, and we randomly sampled source flux from a Euclidean power law distribution with
a minimum flux threshold of 26.5 mag. After inserting fake sources,
we remade our detection image and then performed our standard extraction
procedure. We repeated this simulation 10 times resulting in
  100,000 fake input sources per tile. 

For a given simulation and tile, we use the same $riz$-band detection
image for all bands. Thus, the number of detected fake objects and
the estimated completeness does not vary with bandpass. In Figure
\ref{complete}, we show the recovered fraction of fake sources as
a function of magnitude (solid curves). Our simulations indicate that
our catalogs are $80\%$ complete at a magnitude of $\sim$$24.4$.
After this threshold, our completeness quickly falls off with a $50\%$
completeness limit at $\sim24.7$ mag. 

We also estimate the purity of our sample by inverting all our science
images (multiplying all pixels values by $-1$), forming an inverted
detection image, and then performing our standard extraction procedure
on these images. Assuming that positive noise spikes are as likely
as negative noise spikes, this analysis will estimate how many false
sources are extracted within our catalog. In Figure \ref{complete},
we show the fraction of false sources as a function of magnitude (dashed
curves). We estimate that false sources only become significant at
magnitudes fainter than our estimated $10\%$ completeness limit.

\begin{figure}[!t]
\includegraphics[width=8.5cm]{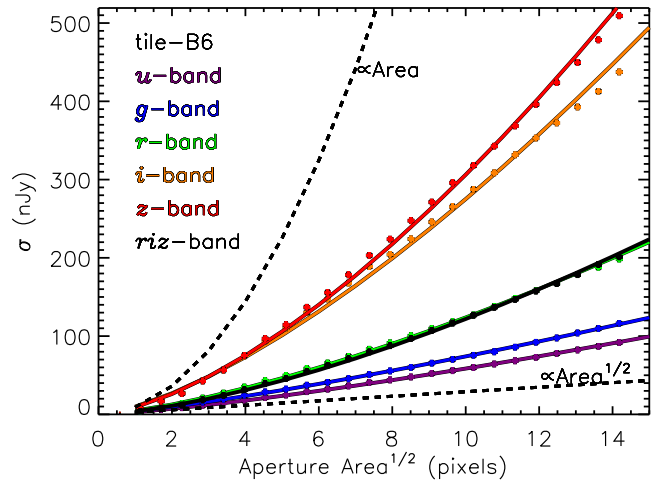}
\caption{Measured aperture noise as a function of the square root of the aperture area
measured in pixels for tile B6. As expected our aperture noise falls between the
dashed lines which correspond to the limiting cases of uncorrelated
and perfectly correlated pixels. We use these aperture noise
measurements to estimate our catalog's
$5\sigma$ depth, see Section \ref{errsect} and Figure \ref{btable}. (The complete figure set (4
images) for all SHELA tiles is available in the online journal.)}

\label{apnoise}

\end{figure}

\begin{deluxetable}{ccccccc} 
\tablecolumns{7} 
\tablewidth{0pc} 
\tablecaption{Survey Area, Exposure Time, and Coefficients for Error Estimates Using Sky Apertures.} 
\tablehead{ 
\colhead{Tile} & \colhead{Filter} & \colhead{Area} & \colhead{Median t$_{exp}$}  & \colhead{$\sigma_{1}$} & \colhead{$\alpha$} & \colhead{$\beta$}\\
\colhead{ } & \colhead{ } & \colhead{(deg$^2$)} & \colhead{(ks)}  & \colhead{(nJy)} & \colhead{ } & \colhead{ }}
\startdata 
B3 & $u$ & 4.850 & 8.401 & 3.81 & 0.72 & 1.36 \\
  & $g$ & 4.856 & 1.159 & 4.38 & 0.88 & 1.31 \\
  & $r$ & 4.856 & 1.000 & 5.15 & 0.78 & 1.42 \\
  & $i$ & 4.856 & 1.289 & 9.69 & 0.63 & 1.47 \\
  & $z$ & 4.856 & 1.590 & 17.08 & 0.48 & 1.51 \\
  & $riz$ & 4.856 & 3.799 & 4.54 & 0.68 & 1.50 \\\hline
\rule{0pt}{4ex}B4 & $u$ & 5.145 & 10.798 & 3.63 & 0.54 & 1.44 \\
  & $g$ & 5.152 & 1.559 & 4.66 & 0.81 & 1.29 \\
  & $r$ & 5.152 & 1.090 & 7.54 & 0.52 & 1.44 \\
  & $i$ & 5.152 & 1.980 & 13.13 & 0.42 & 1.51 \\
  & $z$ & 5.152 & 3.090 & 12.71 & 0.81 & 1.42 \\
  & $riz$ & 5.152 & 6.271 & 5.83 & 0.58 & 1.51 \\\hline
\rule{0pt}{4ex}B5 & $u$ & 4.910 & 11.991 & 3.69 & 0.57 & 1.42 \\
  & $g$ & 4.912 & 1.269 & 4.73 & 0.87 & 1.29 \\
  & $r$ & 4.912 & 0.600 & 7.65 & 0.61 & 1.47 \\
  & $i$ & 4.912 & 0.690 & 11.58 & 0.73 & 1.47 \\
  & $z$ & 4.912 & 1.500 & 12.60 & 0.64 & 1.54 \\
  & $riz$ & 4.912 & 2.690 & 5.60 & 0.64 & 1.53 \\\hline
\rule{0pt}{4ex}B6 & $u$ & 2.567 & 14.399 & 3.48 & 0.83 & 1.31 \\
  & $g$ & 2.570 & 1.449 & 4.39 & 0.93 & 1.26 \\
  & $r$ & 2.570 & 0.780 & 7.19 & 0.65 & 1.42 \\
  & $i$ & 2.570 & 0.290 & 12.72 & 0.78 & 1.44 \\
  & $z$ & 2.570 & 1.200 & 18.12 & 0.50 & 1.53 \\
  & $riz$ & 2.570 & 2.378 & 5.95 & 0.65 & 1.50
\enddata  
\label{signtable}
\end{deluxetable}

\subsection{Detection Limits and Photometric Errors\label{errsect}}

Following the procedure used by \citetalias{papovich16}, we estimate the
photometric depth of our catalogs using two methods. The first method
uses the results from our completeness simulations. In Figure \ref{inout},
we show the median difference between the measured output and known
input magnitude of fake sources as a function of input magnitude for
tile B4. The error bars indicate 1.48 times the median absolute deviation
($\sigma_{{\rm {MAD}}}$). A magnitude error of $\simeq 0.22$ corresponds
to a SNR of 5. Using these binned $\sigma_{{\rm {MAD}}}$ values, we
interpolate to estimate where $\sigma_{{\rm {MAD}}}=0.22$. We report
these $5\sigma$ detection limits based on our simulations in Table
\ref{btable}. In Figure \ref{uncert}, we show $\sigma_{{\rm {MAD}}}$
as a function of magnitude for tile B5 (solid curves). We note that
these depth estimates do not account for systematic errors, but our
simulations show that the median difference between the input and
recovered magnitude is near zero for objects brighter than $\sim24.5$
mag (e.g., see Figure \ref{inout}).

For the second  photometric  depth estimate, we  measured the image background noise as a function of photometric aperture size. We performed aperture
photometry on 10,000 randomly placed sky positions. We ensured that
the apertures did not overlap and that regions containing cataloged objects
were excluded using the segmentation map. We then computed $\sigma_{{\rm {MAD}}}$ for the aperture
fluxes. In Figure \ref{apnoise}, we show $\sigma_{{\rm {MAD}}}$
as a function of the square root of the aperture area for tile B6.
In the limiting case of uncorrelated pixels, $\sigma_{{\rm {MAD}}}$
is proportional to the square root of the aperture area (lower dashed
line). At the other extreme of perfect correlation, $\sigma_{{\rm {MAD}}}$
is proportional to the aperture area (upper dashed line). To estimate
the noise in an aperture of arbitrary size, we follow the
procedure of \citet{gawiser06} and fit our data with a function with the following form:

\begin{figure}[!t]
\includegraphics[width=8.5cm]{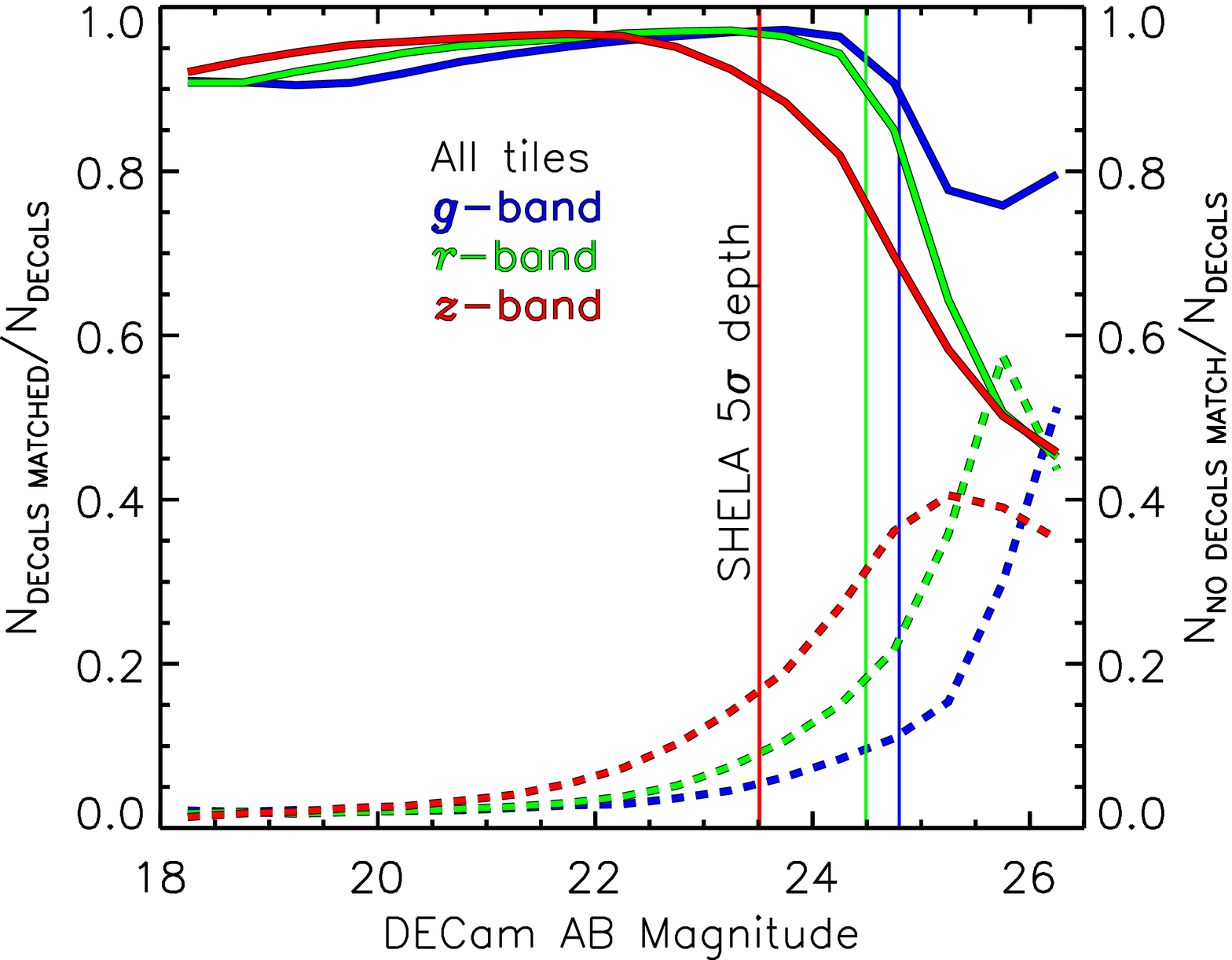}
\caption{The ratio of the number of SHELA sources with DECaLS DR5 counterparts
($N_{{\rm {DECaLS MATCHED}}}$) to the total number of cataloged DECaLS DR5
sources within the SHELA field ($N_{{\rm {DECaLS}}}$; solid red, green, and blue curves)
and the ratio of unmatched SHELA sources ($N_{{\rm {NO DECaLS MATCH}}}$) to the total number of cataloged
DECaLS DR5 sources within the SHELA field ($N_{{\rm {DECaLS}}}$; dashed red, green, and
blue curves) versus magnitude.  We find that our catalog has a
deficit of bright sources ($<21$ mag) relative to DECaLS. Primarily,
we attribute this to artifacts within the DECaLS DR5 catalog (see
Section \ref{verifi}).  
}

\label{complete2}

\end{figure}

\begin{figure}[!t]
\includegraphics[width=8.5cm]{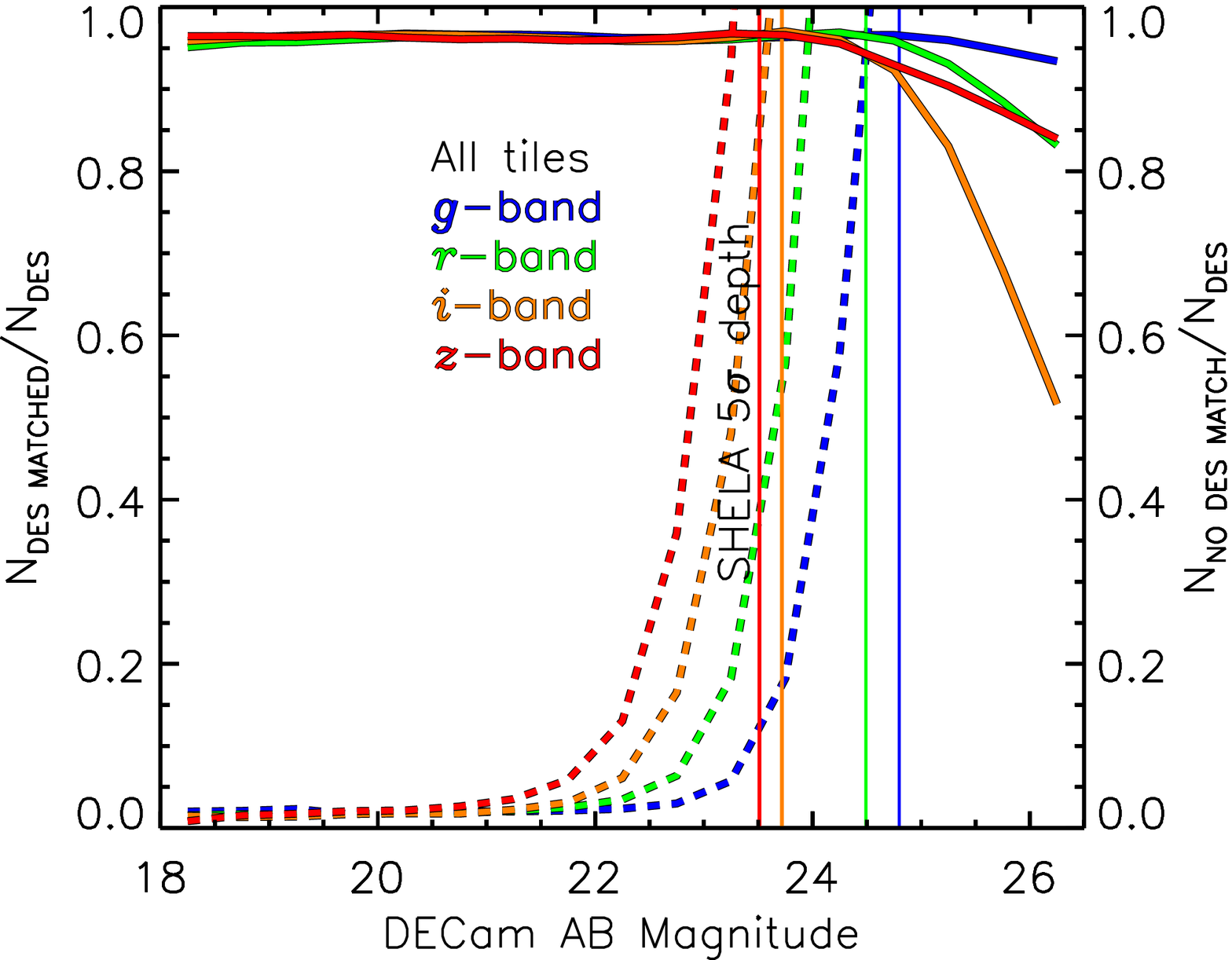}
\caption{The ratio of the number of SHELA sources with DES DR1 counterparts
($N_{{\rm {DES MATCHED}}}$) to the total number of cataloged DES DR1
sources within the SHELA field ($N_{{\rm {DES}}}$; solid red, green, orange, and blue curves)
and the ratio of unmatched SHELA sources ($N_{{\rm {NO DES MATCH}}}$) to the total number of cataloged
DES DR1 sources within the SHELA field ($N_{{\rm {DES}}}$; dashed red, green, orange, and
blue curves) versus magnitude. At bright magnitudes, we find our SExtrator-based
catalog has better agreement with the SExtrator-based DES catalog
relative to the Tractor-based DECaLS catalog.  We attribute the upturn
of unmatched SHELA sources (dashed curves) to DES DR1 incompleteness (see
Section \ref{verifi}).   }

\label{complete3}

\end{figure}

\begin{equation}
\sigma_{{\rm {MAD}}}=\sigma_{1}\alpha A^{\beta/2}\label{eq:10}
\end{equation}

\noindent where $\sigma_{1}$ is the median standard deviation of un-flagged background
pixels, $A$ is the aperture area, $\alpha$ and $\beta$ are free
parameters. We report these values along with the area and median exposure time
of each tile in Table \ref{signtable}. As expected
our best fit $\beta$ values fall between $\beta=1-2$  indicating partially
correlated pixels.

In Figure \ref{uncert}, we show the magnitude uncertainty computed from our B5 PSF apertures
(dashed curves). We have scaled up the PSF aperture uncertainty to
a total uncertainty by dividing by $\sim0.70$. We find that at bright
magnitudes our two methods of estimating photometric uncertainties
are within $0.02$ mag (see solid and dashed curves in Figure \ref{uncert}).
We find that our uncertainty results based on our completeness simulations
are generally higher but can be brought into better agreement by ensuring
that simulated sources do not overlap with real objects. Astrophysically,
there is no reason to prevent sources from overlapping. Thus, we report
$5\sigma$ detection limits based on our simulations (which allow
overlaps) and based on measured fluctuations within sky apertures in Table \ref{btable}.

For our SHELA catalog, we compute photometric errors by inputting
the aperture area used to measure the object into Equation \ref{eq:10}.
For PSF apertures we increase the aperture uncertainty to a total
uncertainty by dividing by a correction factor, $c_{{\rm {psf}}}\sim0.70$
(exact values are listed in Table \ref{btable}). We add this value
in quadrature with an estimated systematic uncertainty of $\sigma_{{\rm {sys}}}=0.05$
mag. This $\sigma_{{\rm {sys}}}$ term accounts for systematic errors
in our photometric zero-points (see Section \ref{szpt}) and in the
offset seen between our two methods of estimating photometric uncertainties
(discussed above). Our cataloged total photometric flux error is:

\begin{equation}
\sigma_{{\rm {tot}}}^{2}=\left(\frac{\sigma_{\rm{MAD}}\
    \rm{rms}_{i}}{c_{{\rm {psf}}}\ \rm{rms}_{{\rm
        {med}}}}\right)^{2}+\left(\frac{ln(10)\ \sigma_{{\rm {sys}}}\ f_{i}}{2.5}\right)^{2}
\end{equation}

\noindent where $\sigma_{{\rm {MAD}}}$ is computed from Equation \ref{eq:10},
rms$_{{\rm {med}}}$ is the median rms for a given tile, rms$_{i}$
is the rms measured at the source's position  ($c_{{\rm {psf}}}=1$
for MAG\_AUTO apertures) and $f_{i}$ is the source's flux density.

\subsection{Comparison of the SHELA Catalog to DECaLS and DES}

\noindent \label{verifi}

As discussed in Section \ref{szpt}, DECaLS has independently reduced
the same DECam $grz$-band data within our survey area. Consistent
with our work, DECaLS DR5 calibrates photometry to the AB DECam filter
system. This provides us with a direct method for verifying our $g$-,
$r$-, and $z$-band catalogs. In Section \ref{szpt}, we used DECaLS
to verify our zero-points. In Figure \ref{complete2}, we use DECaLS to estimate our catalog's
completeness and purity. By making the assumption that DECaLS is `truth',
we estimate the our catalog's completeness by computing the ratio
of the number of SHELA sources with DECaLS DR5 counterparts ($N_{{\rm {MATCHED}}}$)
to the total number of cataloged DECaLS DR5 sources within the SHELA
field ($N_{{\rm {DECaLS}}}$). To minimize the effects of photometric
errors, we exclude all of our cataloged magnitudes with internal SExtractor
flags denoting saturated, truncated, or corrupted pixels (internal
SExtractor flag $>3$). We also exclude magnitudes with non-zero external
SExtractor flags which we use primarily to indicate the close proximity
to a saturated star (see Section \ref{sect_se} for details). To ensure
a fair comparison we also exclude DECaLS sources that fall within
our bright star mask.

For magnitudes brighter than our $\sim5\sigma$ detection limit, we
expect $N_{{\rm {MATCHED}}}\sim N_{{\rm {DECaLS}}}$ indicating a
$\sim100\%$ completeness. In Figure \ref{complete2}, we show that
the DECaLS completeness only reaches $\sim95\%$ even for relatively
bright sources ($18<m<21$ mag). To investigate this further, we looked
for Pan-Starrs PS1 counterparts for all of the bright g-band DECaLS sources ($18<m<21$
mag) missed by our catalog. We find that $\sim65\%$
of these sources are also missing from Pan-Starrs PS1. We randomly selected 10 DECaLS objects missed 
by both PS1 and our catalog and visually inspected both DECaLS and our images. These 
sources appear to be predominately located next to extended galaxies
or saturated stars (but by our comparison setup they are all outside
our bright star mask) and do not appear to be real. We estimate that
$\sim90\%$ of these sources are artifacts within the DECaLS DR5 catalog.
Thus, we find no evidence that our catalog's completeness differs
significantly from our simulation estimates 
which indicate a $\sim99\%$ completeness for
objects with $m\leq21$ mag.

 Performing another visual spot check
on the bright $g$-band objects ($18<m<21$ mag) contained by both
PS1 and DR5 but missed by our catalog, we find that in $\sim70\%$ of
these cases our catalog counts two closely paired objects as one extended
object. This is to be expected given our decision to PSF match to
the bandpass with the worst seeing. Since this is estimated to affect
a small percentage of sources, we make no correction to our catalog. We also note 
that our completeness relative to DECaLS DR5 declines to $\sim90\%$ at our catalog's $5\sigma$ detection limit with the most 
prominent decline in the $z$-band.  Visually spot checking both DECaLS and
our $z$-band images, we estimate that $\sim50\%$ of these  missing
$5\sigma$ sources ($23<m<23.5$ mag) are cosmic rays contained within the DECaLS catalog.  
The remaining $\sim50\%$ appear to be real sources missed by our catalog.

We also use DECaLS to estimate the impurity of our catalog. Assuming
that DECaLS DR5 is `truth', our catalog's impurity is computed by
taking the ratio of SHELA sources with no DECaLS counterparts ($N_{{\rm {NO\:MATCH}}}$)
to the total number of cataloged DECaLS sources within the SHELA field
($N_{{\rm {DECaLS}}}$). Having a large percentage of unmatched objects
could indicate significant artifacts in our catalog. In Figure \ref{complete2},
we show that our catalog's impurity is estimated to be small ($\lesssim5\%$)
at bright magnitudes and increases to $\sim15\%$ at
our $5\sigma$ detection limit. Visually spot checking both DECaLS and
our $z$-band images, we find that the majority of these $\geq5\sigma$
objects  ($m<23.5$ mag) appear to be real in our images, but are not detected in the
 DECaLS DR5 images. Many of the missed objects are
aligned with diffraction spikes from bright stars and it could be that DECaLS DR5
 has applied a more conservative star mask in these regions.

While we have emphasized subtle differences, overall we find that
DECaLS DR5 is in agreement with our catalog and appears to have a
comparable depth (see Table \ref{btable}).  To quantify the depth of DECaLS DR5, we have selected all DR5 objects within our field
 of view and found the median cataloged $5\sigma$ point source
 depth (their `psfdepth' estimate).  These DR5 depths, which are based upon the
 formal errors in the Tractor catalogs for point sources, are
roughly consistent with our SHELA depth estimates.  Specifically,
the depth of DECaLS DR5 is on median 0.1\,mag deeper than our sky aperture depth and 0.3\,mag deeper than our
 simulation detection limit as reported in Table \ref{btable}.

In Figure \ref{complete3}, we replicate Figure \ref{complete2} but use DES DR1 as 
the $griz$-band reference catalog.  We find that our catalogs are in close 
agreement ($\pm5\%$) over a magnitude range of $18-22$ AB.  For fainter magnitudes the 
fraction of unmatched objects increases rapidly, which we interpret as DES incompleteness.    
We estimate that DES DR1 is $90\%$ complete at $g=23.5$, $r=23.0$, $i=22.5$, and $z=22.2$.  
These estimates are within $0.3$ mag of the reported DES DR1 $95\%$ completeness \citep{abbott18}.

\begin{figure}[!t]
\includegraphics[width=8.5cm]{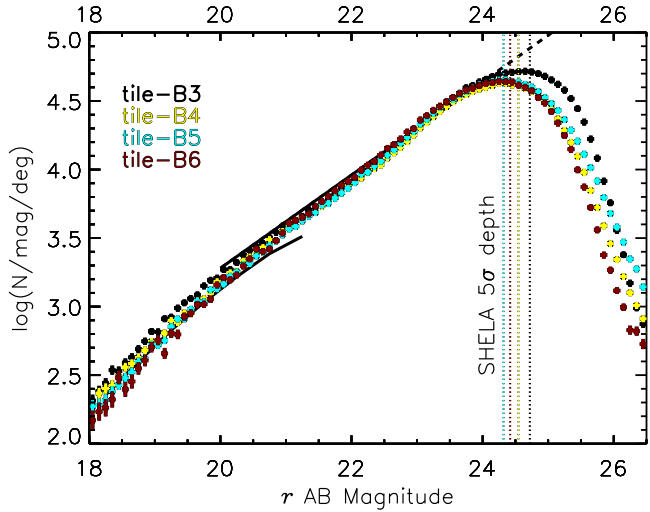}
\caption{$r$-band differential galaxy number counts for each SHELA tile. For
comparison we show the best-fit line ($\log(N)=-3.52+0.34R$) to
the $20<R<24$ $R$-band galaxy counts presented in
\citet[][]{gawiser06}.  Their best-fit line is solid over the
  observed magnitude range and dashed for fainter magnitudes.
We also show bright SDSS $r$-band galaxy counts ($r<21.25$ black curve) from \citet[][]{yasuda01}.
We attribute the downturn in our number counts at $r>24$ to incompleteness
in our catalog. We indicate our catalog's $5\sigma$ depth with
  vertical dotted lines (see Section \ref{errsect} for details)}

\label{counts}

\end{figure}

\subsection{Number Counts}

In Figure \ref{counts}, we show our $r$-band differential galaxy
number counts with Poisson error bars for each SHELA tile. We identified
and removed stars from our sample by cross-referencing our catalog
to known SDSS stars. The reliability of SDSS star classification is
expected to be robust for $r<22$ \citep{lupton01}. For $r>22$, the overall number
counts are dominated by galaxies with stellar contribution estimated
to be $<10\%$. As discussed in Section \ref{verifi}, we remove cataloged
sources with internal SExtractor flags $>3$ and external SExtractor
flags $>0$.

\begin{figure*}[]
\includegraphics[width=17.5cm]{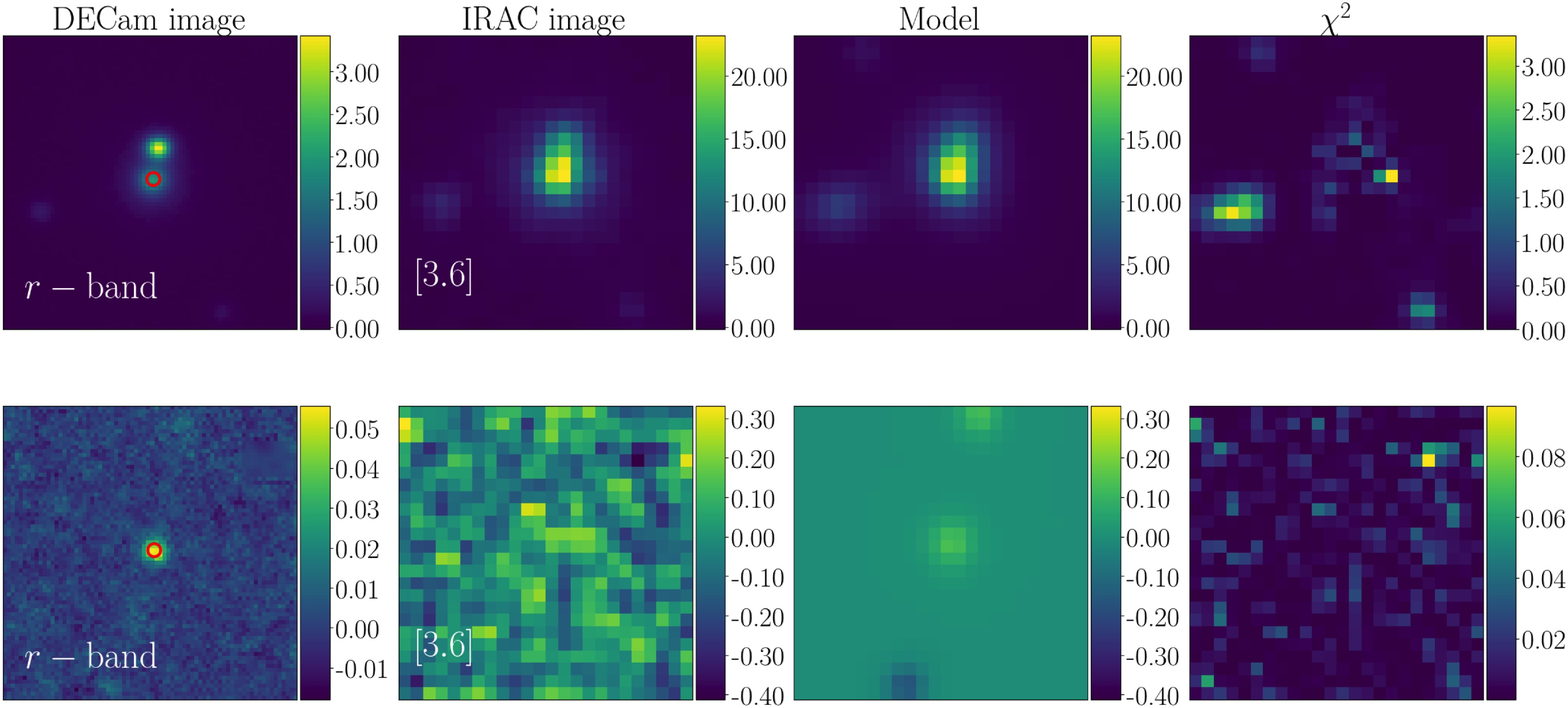}
\caption{\emph{Top}: Example of our forced photometry procedure for a
  source that is clearly blended in the IRAC $3.6\mu$m image but
  resolved in the DECam $r-$band image. The cutout
dimensions are $20''\times20''$ and the source was modeled using a
deVaucouleurs profile. The first two columns show the original images
in DECam $r-$band and IRAC 3.6 $\mu$m band, respectively. The third
column shows the source model convolved with the PRF of each IRAC
band, and the fourth column shows the $\chi^2$ residual maps. The
colorbar units are $\mu$Jy. We indicate the DECam source whose its
IRAC photometry is measured as red circle symbol in the the first
column. \emph{Bottom:} Example of a source with no blending issues,
but that is undetected in the original IRAC-selected catalog \citepalias{papovich16}.
The source was modeled using an exponential profile.}

\label{fig:tractorexample}

\end{figure*}

Our number counts agree with previous studies. In Figure
\ref{counts}, we illustrate this by showing the best-fit line ($\log(N)=-3.52+0.34R$)
to the the $20<R<24$ $R$-band galaxy counts presented in \citet[][]{gawiser06}.
We also show the $r$-band galaxy number counts from SDSS \citep[black curve limited to bright $<21$ mag;][]{yasuda01}.
We attribute the downturn in our number counts at $r>24$ to incompleteness
in our catalog. Our simulations indicate that our incompleteness is
small for $r<23.5$ and declines to $50\%$ at $r=24.5$. Consistent
with our completeness simulations, our number counts indicate that
our B3 tile is marginally deeper than our other tiles.

\section{Spitzer/IRAC Forced Photometry for DECam-selected sources}
\label{tractor}
In this section we describe how we construct a mixed-resolution
muti-band catalog of DECam $ugriz$ + Spitzer/IRAC 3.6 and 4.5 $\mu$m
bands. We generate an accurate multi-band catalog of
DECam $ugriz$ + Spitzer/IRAC 3.6 and 4.5 $\mu$m bands by performing a
``forced photometry with \textit{The Tractor} image modeling code
\citep{lang16a,lang16b}. This technique employs prior measurements of
source positions and surface brightness profiles from a
high-resolution band to model and fit the fluxes of the source in the
remaining bands. We specifically use \textit{The Tractor} to optimize
the likelihood for the photometric properties of DECam sources in
in both the  IRAC 3.6 and 4.5 $\mu$m bands given initial information on the
source obtained from DECam catalog and IRAC image parameters.  Our approach allows us to detect extremely faint sources that fall well below the IRAC $5\sigma$ depth threshold of
  $22$ AB mag \citepalias{papovich16}.  Additionally, the improved
DECam resolution allows us to accurately measure fluxes for blended IRAC
sources.

\subsection{Image Calibration Parameters and Input Catalog}
\par We supply \textit{The Tractor} with the input image
of SHELA IRAC 3.6 and 4.5 $\mu$m images \citepalias{papovich16} with
corresponding image calibration information, including rms maps (noise model), the empirical IRAC point response function (PRFs), and image astrometric information (WCS). In addition, we supply our DECam catalog in each tile as an input catalog of prior source parameters, including source positions, brightness, and surface brightness profile shapes (effective radius, position angle, and axis ratio).
\subsection{Surface Brightness Profile Modeling}
\label{sec:tractormodeling}
\par Broadly speaking, \textit{The Tractor} proceeds by rendering a
model of a galaxy or a point source convolved with the IRAC PRF at each IRAC band and then performs a linear least-square fit for source fluxes such that the sum of source fluxes is closest to the actual image pixels, with respect to the noise model. Finally, the Tractor provides the measurement IRAC flux of each DECam source with the lowest reduced chi squared value ($\chi^2_{\mathrm{red}})$.

\par In practice, we extract IRAC 3.6 and 4.5 $\mu$m image cutouts of
each DECam source. We select a cutout size of ${20}'' \times {20}''$,
which represents a trade-off between minimizing computational costs
related to larger cutout sizes and ensuring that the contributing sources lie well within the cutout extent.

\par For each DECam source, we measure its IRAC fluxes with three
brightness profiles: a point source profile, an exponential profile
(equivalent to a S\'{e}rsic profile with $n=1$), and a deVaucouleurs
profile (equivalent to a S\'{e}rsic profile with $n=4$). In our final
output catalog, we report optimized IRAC fluxes and model profiles ($0=$point source, $1=$exponential, and $4=$deVaucouleurs
profile) that give the lowest reduced chi squared ($\chi_{\mathrm{red}}^2$) value.  Objects with
dubious Tractor photometry are assigned a non-zero Tractor flag.
Specifically, Tractor flag values indicate $5\sigma$ outliers from the
existing SHELA IRAC SExtractor catalog (flag$=1$) and non-optimized
Tractor extraction (flag$=2$).

\par To avoid unphysical results due to optimizing too many sources parameters (i.e., source positions, shapes, and brightness) and the presence of crowded neighboring sources or bright and extended nearby sources, we perform \textit{The Tractor} forced photometry twice for each DECam source of interest. 
We begin with  identifying and masking out bright and extended DECam
sources located within ${20}''$ from a source of interest. This
includes masking out IRAC-only sources that are not detected in our DECam catalog. For the
first optimization, we exclude the source of interest and other
sources within ${4}''$ aperture radius from the modeling.   The
  light profile of these neighboring sources located within a  ${4}''$
  aperture radius could potentially blend with that of a source of
  interest, and therefore we will simultaneously fit these neighboring
  sources and a source of interest during the second optimization.  We
simultaneously model all other neighboring sources located at distance
$>{4}''$ within the cutout by holding all image calibration
parameters, positions, and  surface brightness profile shapes of
neighboring sources fixed except for their brightnesses, which is allowed to be optimized. 

We then subtract the resulting modeled image cutout of neighboring sources from the IRAC image cutout, and are therefore left with the image of the source of interest and its neighboring sources located within ${4}''$  aperture radius.

\par For the second optimization, we run \textit{The Tractor} on the
IRAC ``neighboring sources subtracted'' cutout by holding all image
calibration parameters fixed and allow source position, brightness,
and effective radius (for an exponential profile or a deVaucouleurs
profile) to be optimized. We allow the position of the source of
interest to be varied within 1 arcsecond relative to the input source
position from DECam catalog. In addition, if the modeled effective
radius output from \textit{The Tractor} is unphysical (i.e., a
negative value), we then rerun \textit{The Tractor} by fixing the
effective radius to the input effective radius from DECam
catalog. Examples of the original multi-band images, models, and
$\chi^2$ maps for a blended IRAC source and a non-blended, faint IRAC
source for which \textit{The Tractor} has produced improved IRAC
photometry are shown in Figure~\ref{fig:tractorexample}. We emphasize
that the example of the faint source with a measurement based on our
DECam-selected forced photometry was not detected in the original
SHELA IRAC SExtractor catalog. These examples demonstrate the ability
of  our forced photometry to  detect extremely faint objects and
sources known to be blended in the original SHELA IRAC catalog \citepalias{papovich16}.

\begin{figure*}[!t]

\includegraphics[width=18.5cm]{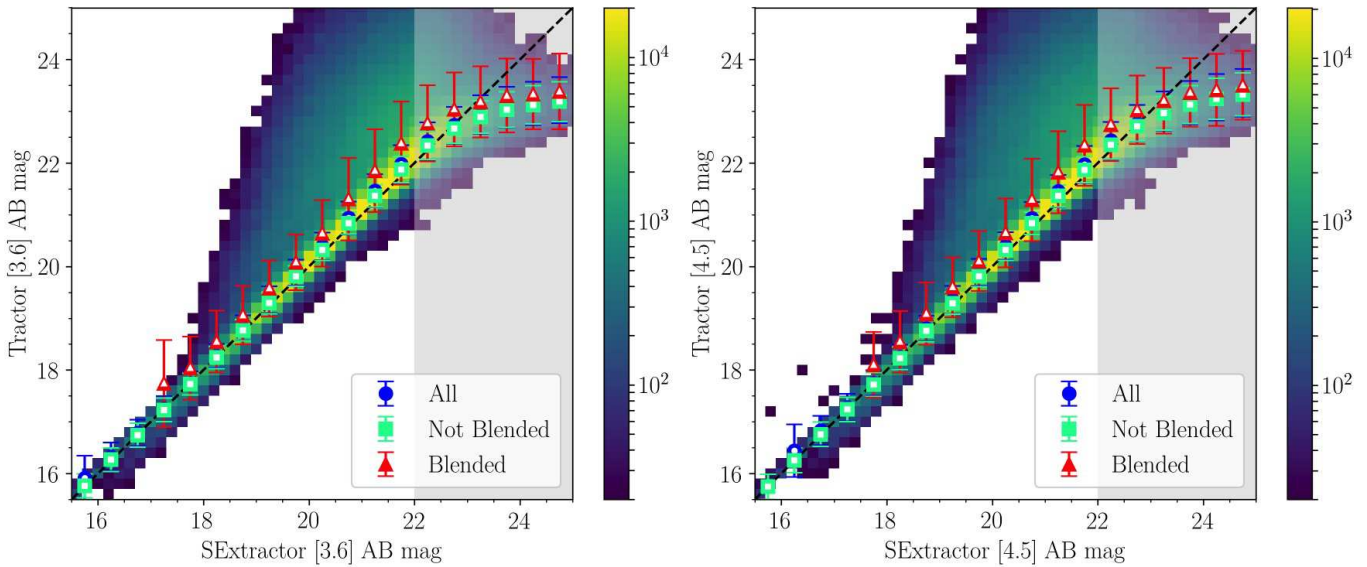}
\caption{The comparison between \textit{The Tractor} and original IRAC
  photometry measured in ${6}''$-diameter apertures, corrected to total
  fluxes. The color scale represents the density of points for all
  sources matched between our DECam catalogs and the IRAC SExtractor
  catalog \citepalias{papovich16}. Data points with error bars show median
  Tractor IRAC magnitudes and corresponding standard deviation from
  median absolute deviations in bins of original IRAC magnitudes for
  all sources (blue circles), non-blended sources (green squares), and
  blended sources (red triangles). Blended IRAC sources are
  identified based on the presence of a nearby source in the DECam
  catalog within ${4}''$. Sources lacking neighbors in DECam within
  ${4}''$ are free of blending issues in IRAC. The dashed line shows
  the one-to-one correspondence between \textit{The Tractor} and
 the \citetalias{papovich16} catalog magnitudes.  The gray-shaded region highlights the
  parameter space below the $80\%$ completeness limit ($22.0$ AB mag).}
\label{fig:tractorfluxcompare}

\end{figure*}

\subsection{Results}
We find that about $47\%$ of the sources in the DECam-selected forced photometry catalogs are best modeled by spatially resolved surface brightness profiles (see Section~\ref{sec:tractormodeling}). Of the resolved sources, the majority are best modeled by an exponential profile ($\sim61\%$) rather than a deVaucouleurs profile ($\sim39\%$).
\par To compare \textit{The Tractor} forced photometry with the
existing SHELA IRAC photometry, we make a catalog of DECam-sources
which are cross-matched with the \citetalias{papovich16} SHELA IRAC
catalog within a search radius of $3''$ (995,607 objects in total). We
identify blended sources in IRAC based on the presence of a nearby
source in the DECam catalog within ${4}''$ (about twice the angular
resolution of the IRAC data), whereas sources lacking neighbors in
DECam within ${4}''$ are free of blending issues in IRAC. Based on
these criteria, we expect at least $35\%$ of the total sources in the DECam input catalog will be blended in the 3.6 and 4.5 $\mu$m IRAC mosaics. This high fraction of blended sources in IRAC is one of the primary motivations for performing forced photometry with \textit{The Tractor}.
\par In Figure~\ref{fig:tractorfluxcompare}, we show a comparison
between the source magnitudes from \textit{The Tractor} forced
photometry and the original IRAC photometry measured in
${6}''$-diameter apertures, corrected to total fluxes
\citepalias{papovich16}. We find that our forced photometry is typically in
good agreement with the SHELA IRAC magnitudes though some scatter is
apparent, particularly for blended sources. The scatter is reduced
when we restrict the comparison to ``isolated'' sources that lack a DECam
neighbor within ${4}''$. The median offsets between the Tractor and
the IRAC aperture-corrected magnitudes are $-0.12,-0.06,0.45$ for
all, not blended, and blended sources brighter than 22 AB, respectively. We expect that
the larger offset and scatter for blended sources are mainly due to
the ability of our \textit{The Tractor} force photometry procedure to
de-blend those sources. The other contributing factors to the scatter
could be spatial PSF variations, inaccurately matched sources in the DECam and IRAC catalogs, and issues with the photometry from the original catalogs. 

\subsubsection{Error Estimates}
\textit{The Tractor} outputs errors on each optimized parameter,
including source brightness. To ensure that \textit{The Tractor}
outputs a photometric error consistent with those from the existing
SHELA IRAC photometric catalog \citepalias{papovich16}, we generate the rms
map for \textit{The Tractor} forced photometry procedure by taking the IRAC 3.6 and 4.5 $\mu$m weight map and scaling $C$ in a $C/\sqrt{(\mathrm{weight~map})}$ image such that the median photometric error output from \textit{The Tractor} are roughly matched with those  measured in ${6}''$-diameter aperture, corrected to total. Here we adopt $C= 0.265$ and 0.287 as scaling factors for our in 3.6 or 4.5 $\mu$m rms images ($C/\sqrt{(\mathrm{weight~map})}$), respectively. Finally, we add the Tractor photometric error in quadrature with an additional error $\sigma_{\mathrm{sys}}=0.05$ mag to account for the median offset between the Tractor and the the aperture-corrected magnitudes for non-blended sources.   The total photometric error $\sigma_{i,c}$ on each DECam source $i$ in IRAC channel $c$ is then given by,

\begin{equation}
\sigma^{2}_{i,c} = \sigma^{2}_{i,c,\mathrm{forced~phot}} + (0.921\sigma_{\mathrm{sys}} \times F_{i,c,\mathrm{forced~phot}})^2
\end{equation}
where $F_{i,c,\mathrm{forced~phot}}$ and  $\sigma_{i,c,\mathrm{forced ~phot}}$ are the flux density and its error measured from \textit{The Tractor} forced photometry procedure.

\subsection{Caveats}
We emphasize that improved photometry of blended IRAC sources can only
be achieved if the blended objects are well resolved in the DECam
catalog. We also note that the accuracy of our photometry with
\textit{The Tractor} will be reduced for highly complex and extended
sources that are not well described by an exponential galaxy profile
or a deVaucouleurs model.  Therefore, we provide a Tractor flag for
each source to indicate failed extractions (flag$=2$) and to indicate where the
Tractor flux differs from the existing SHELA IRAC catalog by more than
$5\sigma$ (flag$=1$):

\begin{equation}
\frac{\left|f_{Tractor}-f_{SExtractor}\right|}{\sqrt{\sigma_{Tractor}^{2}+\sigma_{SExtractor}^{2}}}>5.
\end{equation}

\section{The SHELA Catalog }

In Table \ref{mcat1}, we show a sample of the final SHELA DECam
catalog for tile B3.  Our catalogs include DECam PSF fluxes appropriate for
unresolved sources and SExtractor MAG\_AUTO fluxes for resolved
sources along with ancillary morphology information as described in
Section \ref{sect_se}.  As discussed in Section \ref{verifi}, we
recommend excluding cataloged fluxes with internal SExtractor flag
$>3$ or cataloged fluxes with external SExtractor flag
$>0$\footnote{Given our catalog's convention of rounding to the
  nearest thousandth decimal place for DECam fluxes, there are rare
  cases where the rounded DECam flux error equals ``0.000''.  All
  ``0.000'' flux error sources have non-zero flags indicating
  extraction problems and should be utilized with caution.  To avoid
 null values, we adopt a minimum flux error of ``0.001'' for these sources.}.  We also catalog $J-$ and $K-$band MAG\_AUTO fluxes from
\citet{geach17} for sources with DECam counterparts within
$r=1\farcs5$. This archival near-infrared survey has $5\sigma$
point-source depths of $J=21.4$ and $K=20.9$. DECam sources without
cataloged $J-$ and $K-$band counterparts are assigned a NIR flux of
`$-99$'.  We list 3.6 and 4.5 $\mu$m IRAC Tractor fluxes and ancillary
extraction parameters as discussed in Section
\ref{tractor}. IRAC fluxes with Tractor flags $>1$ indicate
unreliable extractions. Finally, we list EAZY \citep{brammer08}
photometric redshift information which is discused in the following
section.

\subsection{Photometric Redshifts }

\begin{figure}[!t]
\includegraphics[bb=100bp 100bp 620bp 510bp,clip,width=8.5cm]{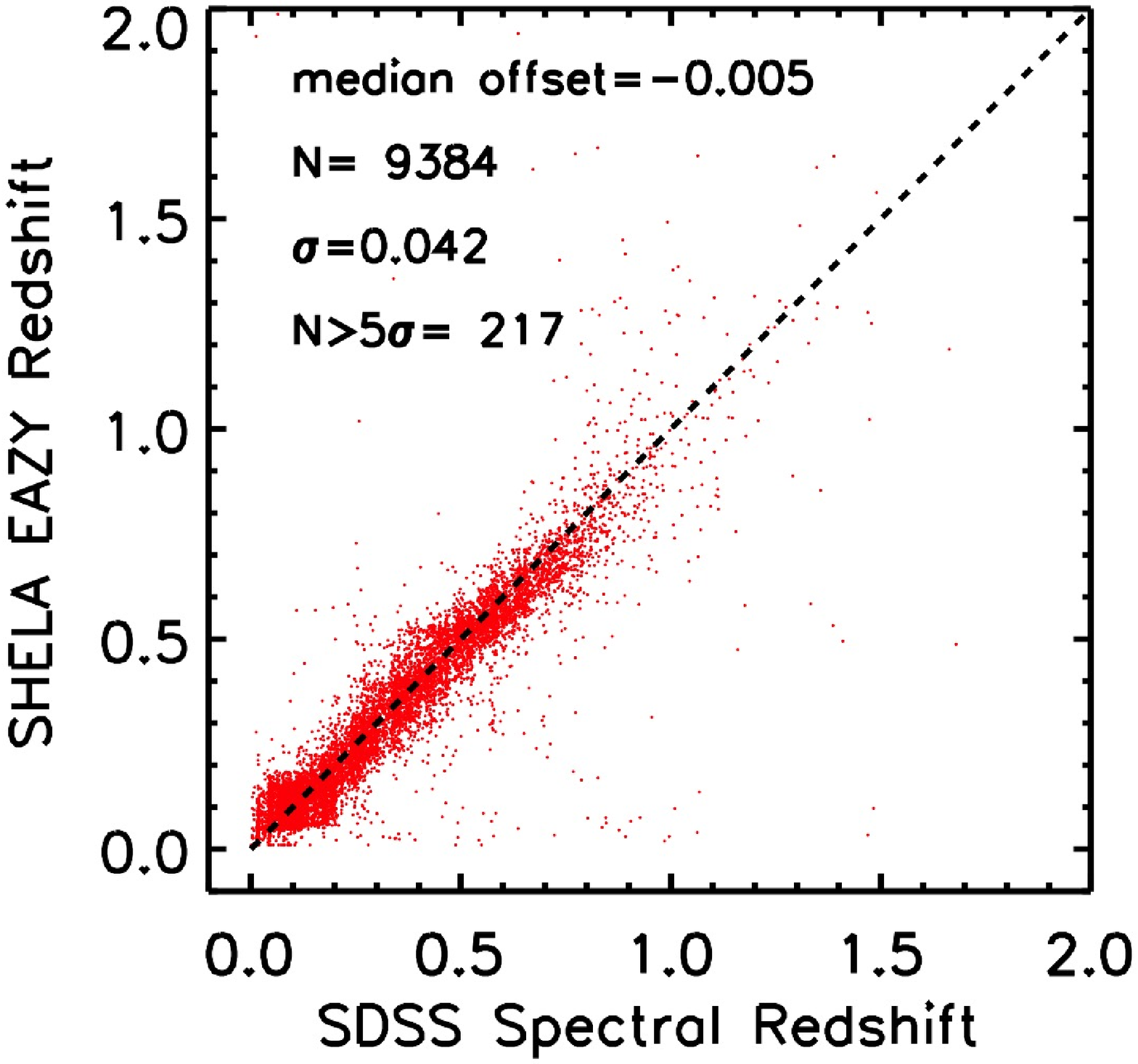}
\includegraphics[bb=100bp 30bp 620bp 260bp,clip,width=8.5cm]{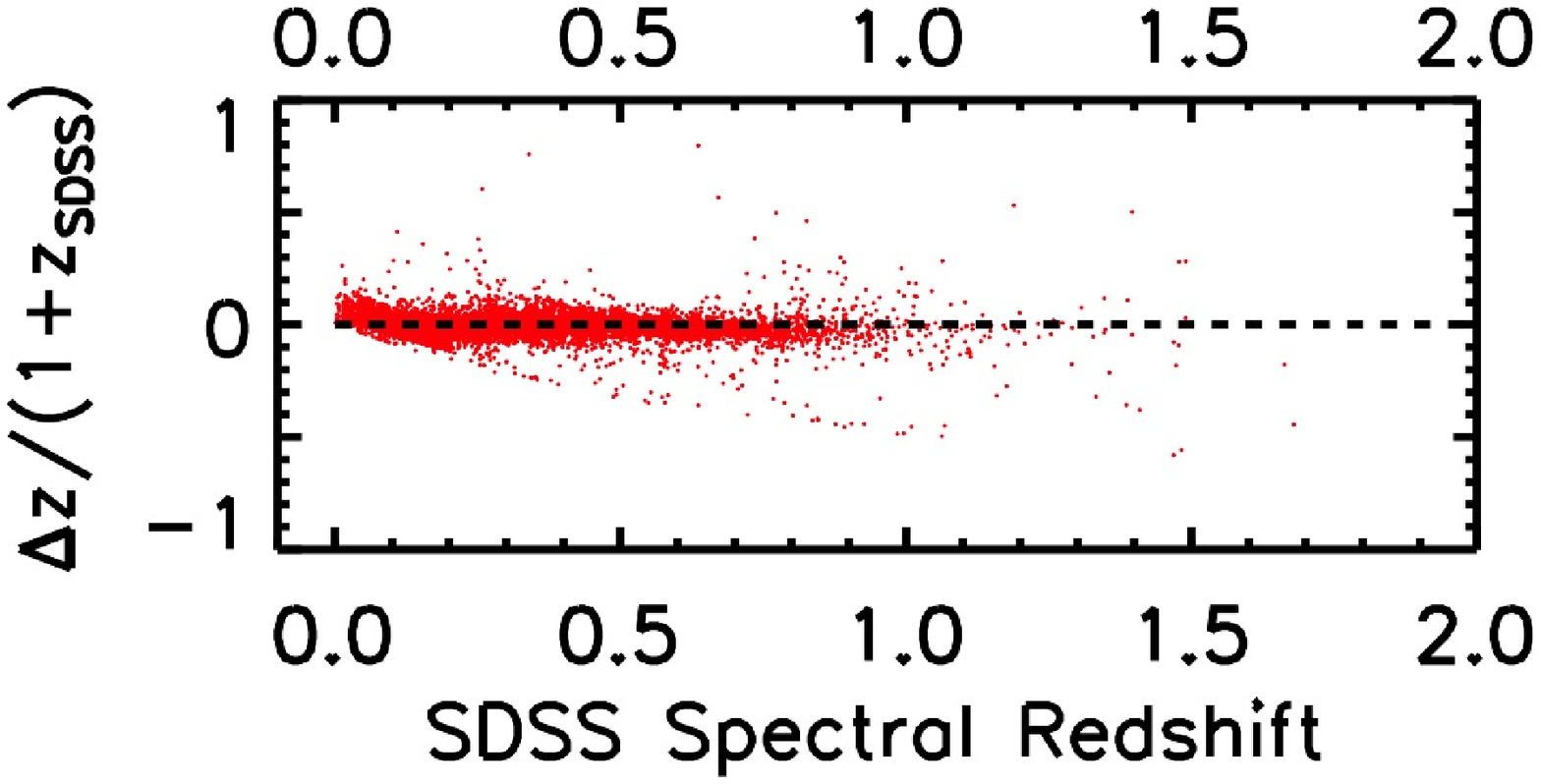}
\caption{Our SHELA photometric $z_{{\rm {peak}}}$  redshifts versus SDSS spectroscopic redshifts.}
\label{photoz}
\end{figure}

\label{sphotoz} In Figure
\ref{photoz} we show our ability to measure photometric redshifts
by comparing our dataset to the available $\left\langle  z
\right\rangle = 0.33$ SDSS spectroscopic
redshifts. In the near future HETDEX will allow us to
extend this analysis out to $z=3.5$ by providing spectroscopic redshifts
for a sample of $N\sim200{,}000$ galaxies at $1.9<z<3.5$. For this work, we match all SDSS $z_{{\rm {spec}}}$ galaxies within
$1''$ of our DECam/IRAC objects. We also utilize the archival $J$ and
$K$-band data from \citet{geach17}. We use SExtractor FLUX\_AUTO fluxes for the $ugrizJK$-bands and use Tractor fluxes for the IRAC bandpasses.

We use the photometric code EAZY \citep{brammer08} to estimate redshifts
from our merged SHELA catalog. We require measured objects to have
5 valid flux measurements, and we utilized EAZY's option to apply a $K$-band magnitude
prior. As discussed in Section \ref{verifi},
we find that the purity of our DECam catalog is improved
by excluding fluxes with internal SExtractor flags $>3$ and external
SExtractor flags $>0$. Additionally, we exclude all IRAC fluxes with
Tractor flags $>1$ which indicate extraction errors.  For the purposes of running EAZY, we assign
these flagged fluxes a large negative number that is below EAZY's
NOT\_OBS\_THRESHOLD.

We display EAZY's $z_{{\rm {peak}}}$ parameter in Figure
\ref{photoz}. This parameter corresponds to the peak probability
of the P($z$) function, and is considered the most accurate  $z_{{\rm {photo}}}$
estimate \citep{muzzin13}. We find  that the median $\Delta z=z_{{\rm {photo}}}-z_{{\rm {spec}}}$
is $-0.005$ and the normalized median absolute deviation, defined as:

\begin{equation}
\sigma_{NMAD}=1.48\times \rm{median}\left(\left|\frac{\Delta z-median(\Delta z)}{1+z_{{\rm {spec}}}}\right|\right),
\end{equation}

\noindent is 0.042 with $2.3\%$ of sources found to be $5\sigma$ outliers.
We find that our photometric results are comparable to those reported
in the EAZY test fields \citep{brammer08}. Here $K$-selected $UBVIJHK+$IRAC
catalogs in CDF-South and other deep fields were found to have median
$\Delta z=-0.005$ and $\sigma_{NMAD}=0.034$ with $5\%$ of the sources
found to be $5\sigma$ outliers for $0<z<6$ galaxies. Future work will improve on our preliminary $z_{{\rm {photo}}}$
estimates by incorporating HETDEX spectroscopic redshifts from $\sim200{,}000$ Ly$\alpha$
emitters at $1.9<z<3.5$ and $\sim200{,}000$ {[}O{\small{}II}{]} emitters at $z<0.5$.

In addition to $z_{{\rm {peak}}}$, we also catalog the EAZY redshift where $\chi^{2}$ is minimized for the
all-template linear combination model without applying a prior
($z_{a}$) and the corresponding minimum $\chi^{2}$ value
($\chi^{2}_{a}$).  Removing the prior lessens the importance of the
$K$-band data and may
produce favorable results for $K$-band faint objects ($K\gtrsim20.9$).

\begin{figure}[!t]
\includegraphics[width=8.5cm]{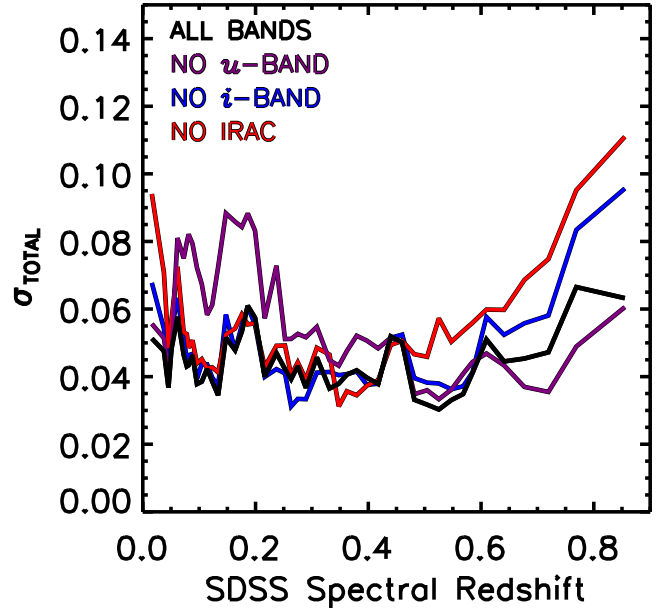}
\caption{Total photometric $z_{{\rm {peak}}}$ redshift error versus SDSS spectroscopic redshift.  }
\label{tsigma}
\end{figure}

Our catalog adds to existing surveys like DECaLS
by providing deep $u-$band and $i-$band measurements along with
IRAC measurements for all detected DECam sources in the SHELA field.
In Figure \ref{tsigma}, we show how the total photometric redshift
error ($\sigma_{{\rm {TOTAL}}}$) is altered if we re-run EAZY with
a DECam bandpass or both IRAC bandpasses excluded from our catalog.
For each EAZY run we keep all input parameters the same with the exception
of the required number of observed bandpasses (N\_MIN\_COLORS) which
we reduce by the number of excluded bandpasses to ensure that we are
able to compare the same sample of $N=9384$ galaxies in each run.
We estimate the total photometric redshift error by adding in quadrature
the random error as measured by $\sigma_{{\rm {NMAD}}}$ with the
systematic error as measured by median$(\Delta z)$. We require each
redshift bin to have $N=200$ galaxies, which restricts our analysis
to $z<1$. Excluding the $u-$band worsens total-sigma
for $z<0.4$ by increasing the $\sigma_{{\rm {NMAD}}}$ term. Excluding
IRAC or to a lesser extent the $i-$band worsens total-sigma for $z>0.4$
by increasing the median$(\Delta z)$ term.   Only by including all
available bandpasses are we able to maintain $\sigma_{{\rm {TOTAL}}}
\sim 0.05$  for our  $z<1$ SDSS sample.

\section{Summary}

SHELA's deep, wide-area multi-wavelength images -- spanning a wavelength
range of 0.35 to 4.5 $\mu$m -- combined with HETDEX's spectroscopic information will enable many
extragalactic studies, including measuring the evolution of
galaxy stellar mass, halo mass, and environment from $1.5<z<3.5$. In this paper, we presented $ugirz$-band DECam catalogs that reach a $5\sigma$
depth of $\sim24.5$ AB mag and cover $17.5$ deg$^{2}$ of the overall
SHELA field. We performed IRAC forced photometry with \textit{The
  Tractor} image modeling code to measure 3.6 and 4.5 $\mu$m  fluxes
for all objects within our DECam catalog.  For our DECam sub-catalog,
we demonstrated consistency with the DECaLS DR5 and the DES DR1.
 We showed that our computed galaxy number counts and photometric redshift
errors are consistent with previous studies.

Ultimately, we plan to tile all of the available
DES $griz$-band data over the full HETDEX fall field, which has a
224 deg$^{2}$ FOV centered at $\alpha$, $\delta=$22.5 deg, 0 deg.
The completed DECam survey for the HETDEX fall field will have tiles
ranging from A1 to C10 and will encompass the presented tiles, B3-B6. 

\medskip{}
\acknowledgements{We thank the anonymous referee for comments that
  substantially improved the manuscript. The authors wish to thank Dustin Lang, John
    Moustakas, Niv Drory, Karl
    Gebhardt, and Rachael Livermore for insightful discussions. IGBW and SLF acknowledge
    support from the National Science Foundation through grants AST
    1518183 and 1614798.  LK and CP acknowledge support from  the
  National Science Foundation through grants AST 1413317 and
  1614668. LK thanks the LSSTC Data Science Fellowship Program, her
  time as a Fellow has benefited this work. This research draws upon data provided by the NOAO Science Archive. NOAO
is operated by the Association of Universities for Research in Astronomy (AURA), Inc. under a cooperative
agreement with the National Science Foundation. This work is based in
part on observations made with the \textit{Spitzer Space Telescope}, which is
operated by the Jet Propulsion Laboratory, California Institute of
Technology under a contract with NASA. The authors acknowledge
  the Texas Advanced Computing Center (TACC) at The University of
  Texas at Austin for providing HPC resources that have contributed to
  the research results reported within this paper. URL:
  http://www.tacc.utexas.edu.

This project used data obtained with the Dark Energy Camera (DECam),
which was constructed by the Dark Energy Survey (DES) collaboration.
Funding for the DES Projects has been provided by 
the U.S. Department of Energy, 
the U.S. National Science Foundation, 
the Ministry of Science and Education of Spain, 
the Science and Technology Facilities Council of the United Kingdom, 
the Higher Education Funding Council for England, 
the National Center for Supercomputing Applications at the University of Illinois at Urbana-Champaign, 
the Kavli Institute of Cosmological Physics at the University of Chicago, 
the Center for Cosmology and Astro-Particle Physics at the Ohio State University, 
the Mitchell Institute for Fundamental Physics and Astronomy at Texas A\&M University, 
Financiadora de Estudos e Projetos, Funda{\c c}{\~a}o Carlos Chagas Filho de Amparo {\`a} Pesquisa do Estado do Rio de Janeiro, 
Conselho Nacional de Desenvolvimento Cient{\'i}fico e Tecnol{\'o}gico and the Minist{\'e}rio da Ci{\^e}ncia, Tecnologia e Inovac{\~a}o, 
the Deutsche Forschungsgemeinschaft, 
and the Collaborating Institutions in the Dark Energy Survey. The Collaborating Institutions are 
Argonne National Laboratory, 
the University of California at Santa Cruz, 
the University of Cambridge, 
Centro de Investigaciones En{\'e}rgeticas, Medioambientales y Tecnol{\'o}gicas-Madrid, 
the University of Chicago, 
University College London, 
the DES-Brazil Consortium, 
the University of Edinburgh, 
the Eidgen{\"o}ssische Technische Hochschule (ETH) Z{\"u}rich, 
Fermi National Accelerator Laboratory, 
the University of Illinois at Urbana-Champaign, 
the Institut de Ci{\`e}ncies de l'Espai (IEEC/CSIC), 
the Institut de F{\'i}sica d'Altes Energies, 
Lawrence Berkeley National Laboratory, 
the Ludwig-Maximilians Universit{\"a}t M{\"u}nchen and the associated Excellence Cluster Universe, 
the University of Michigan, 
{the} National Optical Astronomy Observatory, 
the University of Nottingham, 
the Ohio State University, 
the OzDES Membership Consortium
the University of Pennsylvania, 
the University of Portsmouth, 
SLAC National Accelerator Laboratory, 
Stanford University, 
the University of Sussex, 
and Texas A\&M University.

Based on observations at Cerro Tololo Inter-American Observatory, National Optical
Astronomy Observatory (NOAO Prop. ID 2013B-0438; and PI:
C. Papovich), which is operated by the Association of
Universities for Research in Astronomy (AURA) under a cooperative agreement with the
National Science Foundation.

This project used data from the DECam Legacy Survey. The Legacy Surveys consist of three individual and complementary projects: the Dark Energy Camera Legacy Survey (DECaLS; NOAO Proposal ID 2014B-0404; PIs: David Schlegel and Arjun Dey), the Beijing-Arizona Sky Survey (BASS; NOAO Proposal ID 2015A-0801; PIs: Zhou Xu and Xiaohui Fan), and the Mayall z-band Legacy Survey (MzLS; NOAO Proposal ID 2016A-0453; PI: Arjun Dey). DECaLS, BASS and MzLS together include data obtained, respectively, at the Blanco telescope, Cerro Tololo Inter-American Observatory, National Optical Astronomy Observatory (NOAO); the Bok telescope, Steward Observatory, University of Arizona; and the Mayall telescope, Kitt Peak National Observatory, NOAO. The Legacy Surveys project is honored to be permitted to conduct astronomical research on Iolkam Du'ag (Kitt Peak), a mountain with particular significance to the Tohono O'odham Nation.

This work made use of data from SDSS. Funding for the Sloan Digital Sky Survey IV has been provided by the Alfred P. Sloan Foundation, the U.S. Department of Energy Office of Science, and the Participating Institutions. SDSS-IV acknowledges
support and resources from the Center for High-Performance Computing at
the University of Utah. The SDSS web site is www.sdss.org.

SDSS-IV is managed by the Astrophysical Research Consortium for the 
Participating Institutions of the SDSS Collaboration including the 
Brazilian Participation Group, the Carnegie Institution for Science, 
Carnegie Mellon University, the Chilean Participation Group, the French Participation Group, Harvard-Smithsonian Center for Astrophysics, 
Instituto de Astrof\'isica de Canarias, The Johns Hopkins University, 
Kavli Institute for the Physics and Mathematics of the Universe (IPMU) / 
University of Tokyo, the Korean Participation Group, Lawrence Berkeley National Laboratory, 
Leibniz Institut f\"ur Astrophysik Potsdam (AIP),  
Max-Planck-Institut f\"ur Astronomie (MPIA Heidelberg), 
Max-Planck-Institut f\"ur Astrophysik (MPA Garching), 
Max-Planck-Institut f\"ur Extraterrestrische Physik (MPE), 
National Astronomical Observatories of China, New Mexico State University, 
New York University, University of Notre Dame, 
Observat\'ario Nacional / MCTI, The Ohio State University, 
Pennsylvania State University, Shanghai Astronomical Observatory, 
United Kingdom Participation Group,
Universidad Nacional Aut\'onoma de M\'exico, University of Arizona, 
University of Colorado Boulder, University of Oxford, University of Portsmouth, 
University of Utah, University of Virginia, University of Washington, University of Wisconsin, 
Vanderbilt University, and Yale University.

}

\bibliographystyle{apj} 
\bibliography{wold}
\clearpage \begin{landscape} 
\begin{deluxetable}{cccccccccccccc} 
\tabletypesize{\scriptsize} \setlength{\tabcolsep}{0.02in}
\tablecolumns{14} 
\tablewidth{0pc} 
\tablecaption{SHELA DECam Catalog Sample} 
\tablehead{ 
\colhead{ID} & \colhead{R.A.(J2000)} & \colhead{decl.(J2000)} & \colhead{$a$} & \colhead{$e$} & \colhead{$\theta$} & \colhead{$r_{\rm{Kron}}$} & \colhead{$r_{1/2}$} & \colhead{$f_{u}^{\rm{PSF}}$} & \colhead{$\sigma_{u}^{\rm{PSF}}$}  & \colhead{$f_{u}^{\rm{AUTO}}$} & \colhead{$\sigma_{u}^{\rm{AUTO}}$} & \colhead{Internal} & \colhead{External}\\ 
\colhead{ } & \colhead{(deg)} & \colhead{(deg)} & \colhead{(arcsec)} & \colhead{ } & \colhead{(deg)} & \colhead{ } & \colhead{(arcsec)} & \colhead{($\mu$Jy)} & \colhead{($\mu$Jy)} & \colhead{($\mu$Jy)} & \colhead{($\mu$Jy)}  & \colhead{$u$ Flag}& \colhead{$u$ Flag}\\
\colhead{(1)} & \colhead{(2)} & \colhead{(3)} & \colhead{(4)} & \colhead{(5)} & \colhead{(6)} & \colhead{(7)} & \colhead{(8)} & \colhead{(9)} & \colhead{(10)} & \colhead{(11)} & \colhead{(12)} & \colhead{(13)} & \colhead{(14)}}
\startdata 
B3\_342206 & $ 15.527544$ & $  0.953232$ & $ 0.4$ & $0.11$ & $-78.5$ & $  4.5$ & $  0.8$ & $     0.126$ & $     0.079$ & $     0.139$ & $     0.130$ & $      0$ & $      0$ \\
B3\_342207 & $ 15.527545$ & $ -0.920483$ & $ 2.2$ & $0.17$ & $ -6.2$ & $  3.5$ & $  1.6$ & $     3.467$ & $     0.166$ & $     8.813$ & $     0.598$ & $      2$ & $      0$ \\
B3\_342208 & $ 15.527546$ & $  0.045727$ & $ 1.0$ & $0.43$ & $ 29.3$ & $  5.3$ & $  1.4$ & $     0.374$ & $     0.050$ & $     0.767$ & $     0.213$ & $      3$ & $      0$ \\
B3\_342209 & $ 15.527547$ & $ -0.864834$ & $ 0.5$ & $0.09$ & $-40.9$ & $  6.7$ & $  1.7$ & $     0.084$ & $     0.055$ & $     0.069$ & $     0.174$ & $      3$ & $      0$ \\
B3\_342210 & $ 15.527553$ & $ -0.486625$ & $ 0.3$ & $0.21$ & $-88.8$ & $  8.2$ & $  0.8$ & $     0.285$ & $     0.048$ & $     0.319$ & $     0.077$ & $      0$ & $      0$ \\
B3\_342211 & $ 15.527553$ & $  0.914503$ & $ 0.6$ & $0.14$ & $ 29.3$ & $  3.5$ & $  0.8$ & $     0.421$ & $     0.070$ & $     0.428$ & $     0.105$ & $      0$ & $      0$ \\
B3\_342212 & $ 15.527556$ & $ -0.541192$ & $ 0.8$ & $0.13$ & $ 14.3$ & $  4.3$ & $  1.1$ & $     0.659$ & $     0.059$ & $     1.045$ & $     0.180$ & $      0$ & $      0$ \\
B3\_342213 & $ 15.527563$ & $ -0.312693$ & $ 0.4$ & $0.23$ & $ 58.4$ & $  5.5$ & $  0.7$ & $     0.057$ & $     0.046$ & $     0.067$ & $     0.079$ & $      0$ & $      0$ \\
B3\_342214 & $ 15.527564$ & $ -0.485899$ & $ 0.6$ & $0.14$ & $ 20.9$ & $  6.0$ & $  1.1$ & $     0.431$ & $     0.048$ & $     0.660$ & $     0.160$ & $      0$ & $      0$ \\
B3\_342215 & $ 15.527565$ & $  0.217230$ & $ 0.3$ & $0.29$ & $-75.0$ & $  8.7$ & $  1.2$ & $     0.064$ & $     0.047$ & $     0.040$ & $     0.089$ & $      0$ & $      0$ \\
B3\_342216 & $ 15.527566$ & $  0.452349$ & $ 0.8$ & $0.01$ & $-85.6$ & $  3.5$ & $  0.7$ & $     1.462$ & $     0.087$ & $     1.364$ & $     0.152$ & $      0$ & $      0$ \\
B3\_342217 & $ 15.527570$ & $  0.760209$ & $ 0.4$ & $0.11$ & $ 16.8$ & $  6.9$ & $  1.2$ & $     0.039$ & $     0.054$ & $    -0.042$ & $     0.156$ & $      0$ & $      0$ \\
B3\_342218 & $ 15.527570$ & $  0.148665$ & $ 0.3$ & $0.53$ & $  8.6$ & $  9.1$ & $  0.9$ & $     0.141$ & $     0.045$ & $     0.077$ & $     0.062$ & $      0$ & $      0$ \\
B3\_342219 & $ 15.527574$ & $ -0.484608$ & $ 0.5$ & $0.22$ & $-33.5$ & $  5.3$ & $  0.9$ & $     0.455$ & $     0.049$ & $     0.476$ & $     0.090$ & $      0$ & $      0$ \\
B3\_342220 & $ 15.527574$ & $  0.871303$ & $ 0.7$ & $0.09$ & $ 31.0$ & $  4.1$ & $  0.9$ & $     0.187$ & $     0.048$ & $     0.062$ & $     0.119$ & $      0$ & $      0$ \\
B3\_342221 & $ 15.527574$ & $  0.801912$ & $ 0.6$ & $0.13$ & $-35.7$ & $  4.5$ & $  0.9$ & $     0.172$ & $     0.045$ & $     0.325$ & $     0.118$ & $      0$ & $      0$ \\
B3\_342222 & $ 15.527575$ & $ -0.923096$ & $ 0.4$ & $0.59$ & $ 45.0$ & $  9.3$ & $  1.4$ & $     0.050$ & $     0.050$ & $     0.158$ & $     0.097$ & $      1$ & $      0$ \\
B3\_342223 & $ 15.527581$ & $ -0.587486$ & $ 0.7$ & $0.19$ & $-32.8$ & $  6.7$ & $  2.2$ & $     0.224$ & $     0.051$ & $     0.240$ & $     0.253$ & $      3$ & $      0$ \\
B3\_342224 & $ 15.527582$ & $ -0.459715$ & $ 0.3$ & $0.22$ & $-16.2$ & $  7.3$ & $  0.9$ & $     0.130$ & $     0.050$ & $     0.169$ & $     0.099$ & $      0$ & $      0$ \\
B3\_342225 & $ 15.527587$ & $  0.983389$ & $ 1.1$ & $0.13$ & $ 35.9$ & $  3.7$ & $  1.3$ & $     1.121$ & $     0.112$ & $     2.540$ & $     0.405$ & $      0$ & $      0$ \\
B3\_342226 & $ 15.527598$ & $  0.186228$ & $ 0.7$ & $0.15$ & $-66.3$ & $  3.5$ & $  0.8$ & $     0.584$ & $     0.054$ & $     0.716$ & $     0.099$ & $      0$ & $      0$ \\
B3\_342227 & $ 15.527602$ & $  0.923314$ & $ 0.4$ & $0.12$ & $-84.7$ & $  7.5$ & $  1.3$ & $     0.468$ & $     0.063$ & $     0.590$ & $     0.186$ & $      0$ & $      0$ \\
B3\_342228 & $ 15.527604$ & $  0.230341$ & $ 0.6$ & $0.11$ & $  6.1$ & $  3.9$ & $  0.8$ & $     0.726$ & $     0.058$ & $     0.651$ & $     0.106$ & $      0$ & $      0$ \\
B3\_342229 & $ 15.527604$ & $ -0.855998$ & $ 0.3$ & $0.26$ & $ 20.9$ & $  8.2$ & $  0.7$ & $     0.061$ & $     0.051$ & $    -0.005$ & $     0.077$ & $      0$ & $      0$ \\
B3\_342230 & $ 15.527604$ & $ -0.146012$ & $ 0.3$ & $0.27$ & $-20.3$ & $  7.9$ & $  0.7$ & $     0.075$ & $     0.046$ & $    -0.023$ & $     0.067$ & $      0$ & $      0$ \\
B3\_342231 & $ 15.527611$ & $  0.581601$ & $ 0.5$ & $0.20$ & $ 58.6$ & $  5.2$ & $  0.8$ & $     0.470$ & $     0.062$ & $     0.252$ & $     0.114$ & $      0$ & $      0$ \\
B3\_342232 & $ 15.527616$ & $ -0.157985$ & $ 0.5$ & $0.16$ & $ 44.0$ & $  5.8$ & $  0.9$ & $     0.240$ & $     0.048$ & $     0.326$ & $     0.121$ & $      0$ & $      0$ \\
B3\_342233 & $ 15.527617$ & $  0.386069$ & $ 0.4$ & $0.01$ & $-31.2$ & $  4.4$ & $  0.8$ & $     0.312$ & $     0.049$ & $     0.318$ & $     0.080$ & $      0$ & $      0$ \\
B3\_342234 & $ 15.527618$ & $ -0.708591$ & $ 0.6$ & $0.35$ & $-58.7$ & $  3.5$ & $  0.8$ & $     0.418$ & $     0.054$ & $     0.369$ & $     0.073$ & $      0$ & $      0$ \\
B3\_342235 & $ 15.527623$ & $ -0.015360$ & $ 0.4$ & $0.20$ & $-19.0$ & $  6.5$ & $  0.9$ & $     0.140$ & $     0.045$ & $     0.238$ & $     0.096$ & $      0$ & $      0$ \\
B3\_342236 & $ 15.527623$ & $ -0.849101$ & $ 1.0$ & $0.23$ & $-19.9$ & $  4.1$ & $  1.2$ & $     0.661$ & $     0.055$ & $     1.690$ & $     0.204$ & $      0$ & $      0$ \\
B3\_342237 & $ 15.527628$ & $  0.790137$ & $ 1.0$ & $0.04$ & $ 25.0$ & $  3.5$ & $  1.1$ & $     2.394$ & $     0.120$ & $     3.911$ & $     0.246$ & $      0$ & $      0$ \\
B3\_342238 & $ 15.527630$ & $  0.211697$ & $ 0.7$ & $0.00$ & $ 83.7$ & $  5.3$ & $  1.2$ & $    -0.005$ & $     0.044$ & $    -0.120$ & $     0.165$ & $      3$ & $      0$ \\
B3\_342239 & $ 15.527635$ & $  0.867821$ & $ 0.4$ & $0.18$ & $-64.1$ & $  6.7$ & $  0.9$ & $     0.123$ & $     0.060$ & $     0.050$ & $     0.126$ & $      0$ & $      0$ \\
B3\_342240 & $ 15.527637$ & $  0.109887$ & $ 0.1$ & $0.09$ & $-45.1$ & $  8.9$ & $  0.7$ & $     0.155$ & $     0.045$ & $     0.121$ & $     0.042$ & $      2$ & $      0$ \\
B3\_342241 & $ 15.527639$ & $  0.490189$ & $ 0.7$ & $0.13$ & $ 73.3$ & $  5.8$ & $  1.1$ & $     0.621$ & $     0.053$ & $     0.810$ & $     0.178$ & $      3$ & $      0$ \\
B3\_342242 & $ 15.527645$ & $  0.741641$ & $ 0.5$ & $0.22$ & $  4.7$ & $  7.4$ & $  1.5$ & $     0.151$ & $     0.055$ & $     0.265$ & $     0.180$ & $      1$ & $      0$ \\
B3\_342243 & $ 15.527646$ & $ -1.025758$ & $ 1.3$ & $0.34$ & $ 43.4$ & $  3.5$ & $  1.2$ & $     5.255$ & $     0.251$ & $     9.989$ & $     0.534$ & $      3$ & $      0$ \\
B3\_342244 & $ 15.527649$ & $ -0.386170$ & $ 0.5$ & $0.25$ & $ 60.4$ & $  4.3$ & $  0.8$ & $     0.233$ & $     0.048$ & $     0.325$ & $     0.080$ & $      3$ & $      0$ \\
B3\_342245 & $ 15.527656$ & $ -0.129023$ & $ 1.3$ & $0.43$ & $-12.4$ & $  3.6$ & $  1.1$ & $     0.253$ & $     0.048$ & $     0.550$ & $     0.171$ & $      0$ & $      0$ \\
B3\_342246 & $ 15.527657$ & $ -0.054396$ & $ 0.2$ & $0.41$ & $-18.4$ & $  9.1$ & $  0.7$ & $     0.075$ & $     0.044$ & $     0.097$ & $     0.050$ & $      0$ & $      0$ \\
B3\_342247 & $ 15.527658$ & $  0.506951$ & $ 0.8$ & $0.13$ & $ 25.5$ & $  3.8$ & $  0.9$ & $     1.881$ & $     0.100$ & $     2.138$ & $     0.169$ & $      0$ & $      0$ \\
B3\_342248 & $ 15.527661$ & $ -0.046701$ & $ 1.0$ & $0.18$ & $ 34.4$ & $  3.5$ & $  0.9$ & $     8.493$ & $     0.396$ & $     9.649$ & $     0.481$ & $      0$ & $      0$ \\
B3\_342249 & $ 15.527661$ & $ -0.508430$ & $ 0.8$ & $0.26$ & $ 69.4$ & $  3.5$ & $  0.9$ & $     0.512$ & $     0.050$ & $     0.627$ & $     0.100$ & $      0$ & $      0$ \\
B3\_342250 & $ 15.527662$ & $  0.479859$ & $ 0.5$ & $0.18$ & $-27.8$ & $  7.0$ & $  1.1$ & $     0.071$ & $     0.047$ & $     0.066$ & $     0.137$ & $      0$ & $      0$ \\
B3\_342251 & $ 15.527669$ & $ -0.250235$ & $ 0.2$ & $0.38$ & $  0.4$ & $  9.0$ & $  0.6$ & $     0.082$ & $     0.047$ & $     0.074$ & $     0.047$ & $      0$ & $      0$ \\
\enddata  
\label{mcat1}
\end{deluxetable}
\clearpage
\end{landscape}

\clearpage \begin{landscape} 
\addtocounter{table}{-1}
\begin{deluxetable}{cccccccccccc} 
\tabletypesize{\scriptsize} \setlength{\tabcolsep}{0.02in}
\tablecolumns{12} 
\tablewidth{0pc} 
\tablecaption{SHELA DECam Catalog Sample (continued)} 
\tablehead{ 
\colhead{$f_{g}^{\rm{PSF}}$} & \colhead{$\sigma_{g}^{\rm{PSF}}$}  & \colhead{$f_{g}^{\rm{AUTO}}$} & \colhead{$\sigma_{g}^{\rm{AUTO}}$} & \colhead{Internal} & \colhead{External} & \colhead{$f_{r}^{\rm{PSF}}$} & \colhead{$\sigma_{r}^{\rm{PSF}}$}  & \colhead{$f_{r}^{\rm{AUTO}}$} & \colhead{$\sigma_{r}^{\rm{AUTO}}$} & \colhead{Internal} & \colhead{External}\\ 
\colhead{($\mu$Jy)} & \colhead{($\mu$Jy)} & \colhead{($\mu$Jy)} & \colhead{($\mu$Jy)}  & \colhead{$g$ Flag}& \colhead{$g$ Flag} & \colhead{($\mu$Jy)} & \colhead{($\mu$Jy)} & \colhead{($\mu$Jy)} & \colhead{($\mu$Jy)}  & \colhead{$r$ Flag}& \colhead{$r$ Flag}\\
\colhead{(15)} & \colhead{(16)} & \colhead{(17)} & \colhead{(18)} & \colhead{(19)} & \colhead{(20)} & \colhead{(21)} & \colhead{(22)} & \colhead{(23)} & \colhead{(24)} & \colhead{(25)} & \colhead{(26)}}
\startdata 
$     0.144$ & $     0.077$ & $     0.017$ & $     0.122$ & $      0$ & $      0$ & $     0.515$ & $     0.107$ & $     0.491$ & $     0.181$ & $      0$ & $      0$ \\
$    15.871$ & $     0.734$ & $    38.123$ & $     1.858$ & $      2$ & $      0$ & $    56.910$ & $     2.622$ & $   130.260$ & $     6.062$ & $      2$ & $      0$ \\
$     0.399$ & $     0.063$ & $     0.489$ & $     0.256$ & $      3$ & $      0$ & $     0.490$ & $     0.080$ & $     1.018$ & $     0.382$ & $      3$ & $      0$ \\
$     0.264$ & $     0.069$ & $     0.430$ & $     0.206$ & $      3$ & $      0$ & $     0.559$ & $     0.088$ & $     0.896$ & $     0.292$ & $      3$ & $      0$ \\
$     0.300$ & $     0.070$ & $     0.245$ & $     0.109$ & $      0$ & $      0$ & $     0.380$ & $     0.091$ & $     0.430$ & $     0.153$ & $      0$ & $      0$ \\
$     0.424$ & $     0.068$ & $     0.461$ & $     0.100$ & $      0$ & $      0$ & $     0.689$ & $     0.095$ & $     0.687$ & $     0.148$ & $      0$ & $      0$ \\
$     0.763$ & $     0.071$ & $     1.130$ & $     0.206$ & $      0$ & $      0$ & $     1.352$ & $     0.106$ & $     2.195$ & $     0.333$ & $      0$ & $      0$ \\
$     0.186$ & $     0.066$ & $     0.190$ & $     0.108$ & $      0$ & $      0$ & $     0.454$ & $     0.096$ & $     0.370$ & $     0.168$ & $      0$ & $      0$ \\
$     0.724$ & $     0.071$ & $     1.177$ & $     0.217$ & $      0$ & $      0$ & $     1.104$ & $     0.103$ & $     1.692$ & $     0.353$ & $      0$ & $      0$ \\
$     0.087$ & $     0.059$ & $     0.212$ & $     0.107$ & $      0$ & $      0$ & $     0.322$ & $     0.079$ & $     0.475$ & $     0.155$ & $      0$ & $      0$ \\
$    13.339$ & $     0.618$ & $    12.903$ & $     0.616$ & $      0$ & $      0$ & $    49.914$ & $     2.300$ & $    47.510$ & $     2.200$ & $      0$ & $      0$ \\
$     0.149$ & $     0.059$ & $     0.133$ & $     0.160$ & $      0$ & $      0$ & $     0.311$ & $     0.091$ & $     0.849$ & $     0.282$ & $      0$ & $      0$ \\
$     0.225$ & $     0.058$ & $     0.129$ & $     0.079$ & $      0$ & $      0$ & $     0.280$ & $     0.090$ & $     0.368$ & $     0.131$ & $      0$ & $      0$ \\
$     0.490$ & $     0.066$ & $     0.587$ & $     0.123$ & $      0$ & $      0$ & $     0.683$ & $     0.095$ & $     0.732$ & $     0.189$ & $      0$ & $      0$ \\
$     0.114$ & $     0.061$ & $    -0.052$ & $     0.147$ & $      0$ & $      0$ & $     0.745$ & $     0.085$ & $     0.891$ & $     0.213$ & $      0$ & $      0$ \\
$     0.253$ & $     0.060$ & $     0.441$ & $     0.151$ & $      0$ & $      0$ & $     0.640$ & $     0.083$ & $     0.621$ & $     0.220$ & $      0$ & $      0$ \\
$     0.113$ & $     0.071$ & $     0.263$ & $     0.133$ & $      1$ & $      0$ & $     0.299$ & $     0.083$ & $     0.751$ & $     0.170$ & $      1$ & $      0$ \\
$     0.197$ & $     0.066$ & $     0.517$ & $     0.315$ & $      3$ & $      0$ & $     0.761$ & $     0.089$ & $     2.352$ & $     0.472$ & $      3$ & $      0$ \\
$     0.217$ & $     0.063$ & $     0.224$ & $     0.121$ & $      0$ & $      0$ & $     0.355$ & $     0.079$ & $     0.352$ & $     0.163$ & $      0$ & $      0$ \\
$     1.835$ & $     0.117$ & $     3.818$ & $     0.348$ & $      0$ & $      0$ & $     3.813$ & $     0.211$ & $     8.024$ & $     0.623$ & $      0$ & $      0$ \\
$     0.615$ & $     0.067$ & $     0.623$ & $     0.122$ & $      0$ & $      0$ & $     1.369$ & $     0.106$ & $     1.485$ & $     0.194$ & $      0$ & $      0$ \\
$     0.455$ & $     0.068$ & $     0.757$ & $     0.194$ & $      0$ & $      0$ & $     0.537$ & $     0.093$ & $     1.130$ & $     0.304$ & $      0$ & $      0$ \\
$     0.847$ & $     0.069$ & $     0.941$ & $     0.125$ & $      0$ & $      0$ & $     1.264$ & $     0.100$ & $     1.328$ & $     0.197$ & $      0$ & $      0$ \\
$     0.001$ & $     0.068$ & $    -0.036$ & $     0.100$ & $      0$ & $      0$ & $     0.306$ & $     0.078$ & $     0.259$ & $     0.121$ & $      0$ & $      0$ \\
$     0.065$ & $     0.065$ & $     0.032$ & $     0.091$ & $      0$ & $      0$ & $     0.349$ & $     0.095$ & $     0.345$ & $     0.141$ & $      0$ & $      0$ \\
$     0.441$ & $     0.065$ & $     0.395$ & $     0.117$ & $      0$ & $      0$ & $     0.538$ & $     0.104$ & $     0.616$ & $     0.209$ & $      0$ & $      0$ \\
$     0.276$ & $     0.064$ & $     0.300$ & $     0.153$ & $      0$ & $      0$ & $     0.547$ & $     0.093$ & $     0.655$ & $     0.245$ & $      0$ & $      0$ \\
$     0.373$ & $     0.061$ & $     0.367$ & $     0.097$ & $      0$ & $      0$ & $     0.432$ & $     0.080$ & $     0.482$ & $     0.138$ & $      0$ & $      0$ \\
$     0.626$ & $     0.068$ & $     0.615$ & $     0.089$ & $      0$ & $      0$ & $     0.706$ & $     0.092$ & $     0.797$ & $     0.129$ & $      0$ & $      0$ \\
$     0.242$ & $     0.055$ & $     0.345$ & $     0.113$ & $      0$ & $      0$ & $     0.382$ & $     0.079$ & $     0.399$ & $     0.176$ & $      0$ & $      0$ \\
$     0.960$ & $     0.077$ & $     1.859$ & $     0.255$ & $      0$ & $      0$ & $     1.919$ & $     0.117$ & $     3.195$ & $     0.369$ & $      0$ & $      0$ \\
$     4.083$ & $     0.198$ & $     6.438$ & $     0.364$ & $      0$ & $      0$ & $     7.759$ & $     0.367$ & $    12.786$ & $     0.671$ & $      0$ & $      0$ \\
$     0.122$ & $     0.059$ & $     0.792$ & $     0.209$ & $      3$ & $      0$ & $     0.634$ & $     0.083$ & $     1.554$ & $     0.321$ & $      3$ & $      0$ \\
$     0.067$ & $     0.060$ & $     0.076$ & $     0.123$ & $      0$ & $      0$ & $     0.292$ & $     0.078$ & $     0.281$ & $     0.173$ & $      0$ & $      0$ \\
$     0.236$ & $     0.058$ & $     0.219$ & $     0.054$ & $      2$ & $      0$ & $     0.371$ & $     0.099$ & $     0.320$ & $     0.094$ & $      2$ & $      0$ \\
$     0.744$ & $     0.075$ & $     1.226$ & $     0.254$ & $      3$ & $      0$ & $     1.022$ & $     0.090$ & $     1.691$ & $     0.339$ & $      3$ & $      0$ \\
$     0.237$ & $     0.069$ & $     0.560$ & $     0.215$ & $      1$ & $      0$ & $     0.424$ & $     0.084$ & $     0.986$ & $     0.296$ & $      1$ & $      0$ \\
$    11.033$ & $     0.516$ & $    20.573$ & $     1.012$ & $      3$ & $      0$ & $    18.888$ & $     0.878$ & $    33.499$ & $     1.637$ & $      3$ & $      0$ \\
$     0.323$ & $     0.064$ & $     0.400$ & $     0.104$ & $      3$ & $      0$ & $     0.675$ & $     0.089$ & $     0.644$ & $     0.149$ & $      3$ & $      0$ \\
$     1.523$ & $     0.091$ & $     2.490$ & $     0.226$ & $      0$ & $      0$ & $     5.391$ & $     0.260$ & $     8.575$ & $     0.497$ & $      0$ & $      0$ \\
$     0.102$ & $     0.056$ & $     0.103$ & $     0.061$ & $      0$ & $      0$ & $     0.332$ & $     0.093$ & $     0.289$ & $     0.106$ & $      0$ & $      0$ \\
$     2.035$ & $     0.109$ & $     2.228$ & $     0.179$ & $      0$ & $      0$ & $     2.510$ & $     0.155$ & $     2.926$ & $     0.329$ & $      0$ & $      0$ \\
$    15.220$ & $     0.704$ & $    18.248$ & $     0.860$ & $      0$ & $      0$ & $    27.002$ & $     1.246$ & $    31.261$ & $     1.468$ & $      0$ & $      0$ \\
$     0.755$ & $     0.067$ & $     0.848$ & $     0.126$ & $      0$ & $      0$ & $     1.640$ & $     0.108$ & $     1.904$ & $     0.198$ & $      0$ & $      0$ \\
$     0.161$ & $     0.060$ & $     0.209$ & $     0.167$ & $      0$ & $      0$ & $     0.508$ & $     0.081$ & $     0.718$ & $     0.244$ & $      0$ & $      0$ \\
$     0.113$ & $     0.062$ & $     0.084$ & $     0.062$ & $      0$ & $      0$ & $     0.358$ & $     0.083$ & $     0.297$ & $     0.085$ & $      0$ & $      0$ \\
\enddata  
\label{mcat2}
\end{deluxetable}
\clearpage
\end{landscape}

\clearpage \begin{landscape} 
\addtocounter{table}{-1}
\begin{deluxetable}{ccccccccccccccc} 
\tabletypesize{\scriptsize} \setlength{\tabcolsep}{0.02in}
\tablecolumns{15} 
\tablewidth{0pc} 
\tablecaption{SHELA DECam Catalog Sample (continued)} 
\tablehead{ 
\colhead{$f_{i}^{\rm{PSF}}$} & \colhead{$\sigma_{i}^{\rm{PSF}}$}  & \colhead{$f_{i}^{\rm{AUTO}}$} & \colhead{$\sigma_{i}^{\rm{AUTO}}$} & \colhead{Internal} & \colhead{External} & \colhead{$f_{z}^{\rm{PSF}}$} & \colhead{$\sigma_{z}^{\rm{PSF}}$}  & \colhead{$f_{z}^{\rm{AUTO}}$} & \colhead{$\sigma_{z}^{\rm{AUTO}}$} & \colhead{Internal} & \colhead{External} & \colhead{VICS82ID} & \colhead{$f_{J}^{\rm{AUTO}}$} & \colhead{$\sigma_{J}^{\rm{AUTO}}$}\\ 
\colhead{($\mu$Jy)} & \colhead{($\mu$Jy)} & \colhead{($\mu$Jy)} & \colhead{($\mu$Jy)}  & \colhead{$i$ Flag}& \colhead{$i$ Flag} & \colhead{($\mu$Jy)} & \colhead{($\mu$Jy)} & \colhead{($\mu$Jy)} & \colhead{($\mu$Jy)}  & \colhead{$z$ Flag} & \colhead{$z$ Flag} & \colhead{Geach+17} & \colhead{($\mu$Jy)} & \colhead{($\mu$Jy)}\\
\colhead{(27)} & \colhead{(28)} & \colhead{(29)} & \colhead{(30)} & \colhead{(31)} & \colhead{(32)} & \colhead{(33)} & \colhead{(34)} & \colhead{(35)} & \colhead{(36)} & \colhead{(37)} & \colhead{(38)} & \colhead{(39)} & \colhead{(40)} & \colhead{(41)}}
\startdata 
$     0.823$ & $     0.149$ & $     0.654$ & $     0.258$ & $      0$ & $      0$ & $     1.818$ & $     0.212$ & $     2.272$ & $     0.369$ & $      0$ & $      0$ & VICS82J010206.62+005712.0      & $    -99.00$ & $    -99.00$ \\
$    91.047$ & $     4.195$ & $   203.426$ & $     9.471$ & $      2$ & $      0$ & $   119.496$ & $     5.506$ & $   265.109$ & $    12.401$ & $      2$ & $      0$ & VICS82J010206.61-005513.7      & $    367.53$ & $      1.58$ \\
$     1.175$ & $     0.130$ & $     2.516$ & $     0.628$ & $      3$ & $      0$ & $     1.847$ & $     0.197$ & $     2.917$ & $     0.987$ & $      3$ & $      0$ & VICS82J010206.54+000245.7      & $      3.13$ & $      0.57$ \\
$     1.106$ & $     0.130$ & $     1.617$ & $     0.436$ & $      3$ & $      0$ & $     1.285$ & $     0.210$ & $     1.576$ & $     0.760$ & $      3$ & $      0$ & \nodata & $    -99.00$ & $    -99.00$ \\
$     0.345$ & $     0.147$ & $     0.288$ & $     0.256$ & $      0$ & $      0$ & $     0.565$ & $     0.203$ & $     0.598$ & $     0.364$ & $      0$ & $      0$ & \nodata & $    -99.00$ & $    -99.00$ \\
$     1.349$ & $     0.140$ & $     1.423$ & $     0.216$ & $      0$ & $      0$ & $     2.110$ & $     0.195$ & $     2.094$ & $     0.301$ & $      0$ & $      0$ & VICS82J010206.59+005452.4      & $      8.86$ & $      4.04$ \\
$     2.957$ & $     0.181$ & $     3.891$ & $     0.500$ & $      0$ & $      0$ & $     3.829$ & $     0.267$ & $     5.345$ & $     0.859$ & $      0$ & $      0$ & \nodata & $    -99.00$ & $    -99.00$ \\
$     0.532$ & $     0.132$ & $     0.641$ & $     0.241$ & $      0$ & $      0$ & $     0.863$ & $     0.243$ & $     0.806$ & $     0.453$ & $      0$ & $      0$ & \nodata & $    -99.00$ & $    -99.00$ \\
$     1.060$ & $     0.154$ & $     1.679$ & $     0.601$ & $      0$ & $      0$ & $     1.439$ & $     0.212$ & $     1.685$ & $     0.866$ & $      0$ & $      0$ & \nodata & $    -99.00$ & $    -99.00$ \\
$     0.413$ & $     0.119$ & $     0.941$ & $     0.247$ & $      0$ & $      0$ & $     0.366$ & $     0.178$ & $     0.633$ & $     0.377$ & $      0$ & $      0$ & VICS82J010206.70+001302.0      & $    -99.00$ & $    -99.00$ \\
$   148.397$ & $     6.835$ & $   143.257$ & $     6.606$ & $      0$ & $      0$ & $   226.570$ & $    10.436$ & $   218.371$ & $    10.075$ & $      0$ & $      0$ & VICS82J010206.60+002708.8      & $    301.48$ & $      1.83$ \\
$     0.978$ & $     0.146$ & $     0.992$ & $     0.455$ & $      0$ & $      0$ & $     2.090$ & $     0.195$ & $     2.042$ & $     0.582$ & $      0$ & $      0$ & VICS82J010206.61+004536.5      & $    -99.00$ & $    -99.00$ \\
$     0.540$ & $     0.133$ & $     0.546$ & $     0.197$ & $      0$ & $      0$ & $     0.413$ & $     0.182$ & $     0.305$ & $     0.278$ & $      0$ & $      0$ & \nodata & $    -99.00$ & $    -99.00$ \\
$     0.398$ & $     0.147$ & $     0.570$ & $     0.317$ & $      0$ & $      0$ & $     0.339$ & $     0.174$ & $     0.097$ & $     0.387$ & $      0$ & $      0$ & \nodata & $    -99.00$ & $    -99.00$ \\
$     2.616$ & $     0.166$ & $     3.230$ & $     0.355$ & $      0$ & $      0$ & $     4.423$ & $     0.265$ & $     5.056$ & $     0.549$ & $      0$ & $      0$ & VICS82J010206.62+005216.7      & $      8.89$ & $      1.55$ \\
$     1.635$ & $     0.138$ & $     1.923$ & $     0.355$ & $      0$ & $      0$ & $     3.595$ & $     0.239$ & $     5.010$ & $     0.578$ & $      0$ & $      0$ & VICS82J010206.63+004806.9      & $    -99.00$ & $    -99.00$ \\
$     0.382$ & $     0.120$ & $     0.981$ & $     0.255$ & $      1$ & $      0$ & $     0.819$ & $     0.177$ & $     1.770$ & $     0.384$ & $      1$ & $      0$ & \nodata & $    -99.00$ & $    -99.00$ \\
$     0.877$ & $     0.126$ & $     3.092$ & $     0.733$ & $      3$ & $      0$ & $     0.954$ & $     0.226$ & $     3.431$ & $     1.425$ & $      3$ & $      0$ & \nodata & $    -99.00$ & $    -99.00$ \\
$     0.448$ & $     0.121$ & $     0.606$ & $     0.262$ & $      0$ & $      0$ & $     0.767$ & $     0.242$ & $     0.561$ & $     0.540$ & $      0$ & $      0$ & \nodata & $    -99.00$ & $    -99.00$ \\
$     4.531$ & $     0.313$ & $     8.712$ & $     1.129$ & $      0$ & $      0$ & $     5.813$ & $     0.359$ & $    10.852$ & $     1.241$ & $      0$ & $      0$ & VICS82J010206.63+005900.5      & $    -99.00$ & $    -99.00$ \\
$     2.483$ & $     0.168$ & $     2.828$ & $     0.300$ & $      0$ & $      0$ & $     3.063$ & $     0.271$ & $     3.295$ & $     0.545$ & $      0$ & $      0$ & VICS82J010206.62+001110.5      & $      3.74$ & $      0.69$ \\
$     0.461$ & $     0.140$ & $     0.864$ & $     0.488$ & $      0$ & $      0$ & $     0.574$ & $     0.171$ & $     0.218$ & $     0.617$ & $      0$ & $      0$ & \nodata & $    -99.00$ & $    -99.00$ \\
$     2.097$ & $     0.152$ & $     2.179$ & $     0.298$ & $      0$ & $      0$ & $     2.839$ & $     0.220$ & $     3.164$ & $     0.460$ & $      0$ & $      0$ & VICS82J010206.59+001349.4      & $      1.59$ & $      0.41$ \\
$     0.428$ & $     0.121$ & $     0.455$ & $     0.194$ & $      0$ & $      0$ & $     0.161$ & $     0.201$ & $    -0.014$ & $     0.333$ & $      0$ & $      0$ & \nodata & $    -99.00$ & $    -99.00$ \\
$     0.458$ & $     0.128$ & $     0.277$ & $     0.195$ & $      0$ & $      0$ & $     0.410$ & $     0.232$ & $     0.530$ & $     0.363$ & $      0$ & $      0$ & \nodata & $    -99.00$ & $    -99.00$ \\
$     0.732$ & $     0.135$ & $     0.676$ & $     0.278$ & $      0$ & $      0$ & $     0.862$ & $     0.182$ & $     0.439$ & $     0.386$ & $      0$ & $      0$ & \nodata & $    -99.00$ & $    -99.00$ \\
$     0.937$ & $     0.135$ & $     1.262$ & $     0.370$ & $      0$ & $      0$ & $     1.657$ & $     0.211$ & $     1.833$ & $     0.586$ & $      0$ & $      0$ & \nodata & $    -99.00$ & $    -99.00$ \\
$     0.846$ & $     0.124$ & $     0.891$ & $     0.218$ & $      0$ & $      0$ & $     0.943$ & $     0.182$ & $     1.121$ & $     0.334$ & $      0$ & $      0$ & \nodata & $    -99.00$ & $    -99.00$ \\
$     0.666$ & $     0.123$ & $     0.634$ & $     0.179$ & $      0$ & $      0$ & $     0.629$ & $     0.176$ & $     0.659$ & $     0.264$ & $      0$ & $      0$ & \nodata & $    -99.00$ & $    -99.00$ \\
$     0.491$ & $     0.140$ & $     0.758$ & $     0.329$ & $      0$ & $      0$ & $     0.936$ & $     0.178$ & $     0.790$ & $     0.424$ & $      0$ & $      0$ & \nodata & $    -99.00$ & $    -99.00$ \\
$     4.270$ & $     0.230$ & $     7.930$ & $     0.671$ & $      0$ & $      0$ & $     6.871$ & $     0.361$ & $    11.901$ & $     1.019$ & $      0$ & $      0$ & VICS82J010206.64-005056.5      & $     20.60$ & $      0.98$ \\
$     9.555$ & $     0.455$ & $    16.167$ & $     0.884$ & $      0$ & $      0$ & $    10.931$ & $     0.542$ & $    18.510$ & $     1.209$ & $      0$ & $      0$ & VICS82J010206.61+004724.1      & $     10.67$ & $      1.26$ \\
$     3.051$ & $     0.183$ & $     4.207$ & $     0.543$ & $      3$ & $      0$ & $     6.861$ & $     0.362$ & $     7.645$ & $     0.874$ & $      3$ & $      0$ & VICS82J010206.63+001242.2      & $     11.71$ & $      0.71$ \\
$     0.587$ & $     0.117$ & $     0.785$ & $     0.267$ & $      0$ & $      0$ & $     0.729$ & $     0.172$ & $     0.895$ & $     0.408$ & $      0$ & $      0$ & \nodata & $    -99.00$ & $    -99.00$ \\
$     0.267$ & $     0.130$ & $     0.242$ & $     0.125$ & $      2$ & $      0$ & $     0.292$ & $     0.178$ & $     0.178$ & $     0.173$ & $      2$ & $      0$ & \nodata & $    -99.00$ & $    -99.00$ \\
$     2.081$ & $     0.169$ & $     2.989$ & $     0.648$ & $      3$ & $      0$ & $     2.669$ & $     0.242$ & $     3.636$ & $     1.008$ & $      3$ & $      0$ & \nodata & $    -99.00$ & $    -99.00$ \\
$     0.707$ & $     0.135$ & $     1.280$ & $     0.498$ & $      1$ & $      0$ & $     0.471$ & $     0.179$ & $     0.943$ & $     0.702$ & $      1$ & $      0$ & \nodata & $    -99.00$ & $    -99.00$ \\
$    26.141$ & $     1.217$ & $    45.376$ & $     2.255$ & $      3$ & $      0$ & $    28.413$ & $     1.436$ & $    50.483$ & $     3.757$ & $      3$ & $      0$ & VICS82J010206.64-010132.5      & $     59.93$ & $      1.76$ \\
$     0.940$ & $     0.138$ & $     1.034$ & $     0.242$ & $      3$ & $      0$ & $     1.219$ & $     0.182$ & $     1.314$ & $     0.327$ & $      3$ & $      0$ & \nodata & $    -99.00$ & $    -99.00$ \\
$    11.234$ & $     0.530$ & $    18.388$ & $     0.974$ & $      0$ & $      0$ & $    14.912$ & $     0.714$ & $    24.075$ & $     1.395$ & $      0$ & $      0$ & VICS82J010206.64-000745.2      & $     16.91$ & $      0.45$ \\
$     0.341$ & $     0.128$ & $     0.276$ & $     0.149$ & $      0$ & $      0$ & $     0.452$ & $     0.171$ & $     0.412$ & $     0.201$ & $      0$ & $      0$ & \nodata & $    -99.00$ & $    -99.00$ \\
$     3.768$ & $     0.217$ & $     4.609$ & $     0.451$ & $      0$ & $      0$ & $     4.707$ & $     0.280$ & $     6.327$ & $     0.634$ & $      0$ & $      0$ & VICS82J010206.62+003025.5      & $      6.15$ & $      0.84$ \\
$    38.803$ & $     1.791$ & $    44.759$ & $     2.102$ & $      0$ & $      0$ & $    40.377$ & $     1.870$ & $    46.205$ & $     2.248$ & $      0$ & $      0$ & VICS82J010206.63-000247.9      & $     57.53$ & $      0.57$ \\
$     3.213$ & $     0.196$ & $     3.654$ & $     0.352$ & $      0$ & $      0$ & $     3.514$ & $     0.237$ & $     3.656$ & $     0.461$ & $      0$ & $      0$ & VICS82J010206.59-003030.2      & $      3.27$ & $      0.44$ \\
$     0.516$ & $     0.133$ & $     0.876$ & $     0.431$ & $      0$ & $      0$ & $     0.443$ & $     0.182$ & $     0.948$ & $     0.622$ & $      0$ & $      0$ & \nodata & $    -99.00$ & $    -99.00$ \\
$     0.141$ & $     0.119$ & $     0.100$ & $     0.124$ & $      0$ & $      0$ & $    -0.001$ & $     0.173$ & $    -0.051$ & $     0.182$ & $      0$ & $      0$ & VICS82J010206.68-001502.0      & $    -99.00$ & $    -99.00$ \\
\enddata  
\label{mcat3}
\end{deluxetable}
\clearpage
\end{landscape}

\clearpage \begin{landscape} 
\addtocounter{table}{-1}
\begin{deluxetable}{ccccccccccccccccc} 
\tabletypesize{\scriptsize} \setlength{\tabcolsep}{0.02in}
\tablecolumns{17} 
\tablewidth{0pc} 
\tablecaption{SHELA DECam Catalog Sample (continued)} 
\tablehead{ 
\colhead{$f_{K}^{\rm{AUTO}}$} & \colhead{$\sigma_{K}^{\rm{AUTO}}$} & \colhead{$f_{3.6}$} & \colhead{$\sigma_{3.6}$}  & \colhead{model$_{3.6}$} & \colhead{log($P_{3.6}$)} & \colhead{Tractor} & \colhead{$f_{4.5}$} & \colhead{$\sigma_{4.5}$}  & \colhead{model$_{4.5}$} & \colhead{log($P_{4.5}$)} & \colhead{Tractor} & \colhead{$z_{\rm{spec}}$}  & \colhead{$z_{\rm{peak}}$} & \colhead{$z_{\rm{a}}$}  & \colhead{$\chi^{2}_{\rm{a}}$} & \colhead{$N_{\rm{filters}}$} \\ 
\colhead{($\mu$Jy)} & \colhead{($\mu$Jy)} & \colhead{($\mu$Jy)} & \colhead{($\mu$Jy)} & \colhead{ } & \colhead{ }  & \colhead{$3.6$ Flag}& \colhead{($\mu$Jy)} & \colhead{($\mu$Jy)} & \colhead{ } & \colhead{ } & \colhead{$4.5$ Flag}  & \colhead{SDSS} & \colhead{EAZY} & \colhead{EAZY} & \colhead{EAZY} & \colhead{EAZY}\\
\colhead{(42)} & \colhead{(43)} & \colhead{(44)} & \colhead{(45)} & \colhead{(46)} & \colhead{(47)} & \colhead{(48)} & \colhead{(49)} & \colhead{(50)} & \colhead{(51)} & \colhead{(52)} & \colhead{(53)} & \colhead{(54)} & \colhead{(55)} & \colhead{(56)} & \colhead{(57)} & \colhead{(58)}}
\startdata 
$      9.66$ & $      1.77$ & $     28.76$ & $      1.97$ & $           0$ & $     -2.05$ & $           0$ & $     38.00$ & $      2.46$ & $           0$ & $     -2.72$ & $           0$ & $    -99.00$ & $      1.45$ & $      5.56$ & $      7.19$ & $           8$ \\
$    527.06$ & $      2.94$ & $    313.92$ & $     15.44$ & $           4$ & $     -7.04$ & $           1$ & $    256.42$ & $     12.58$ & $           4$ & $     -8.89$ & $           1$ & $    -99.00$ & $      0.31$ & $      0.40$ & $      5.07$ & $           9$ \\
$      6.86$ & $      1.41$ & $     11.00$ & $      0.76$ & $           4$ & $     -2.82$ & $           0$ & $      8.68$ & $      1.44$ & $           4$ & $     -3.85$ & $           0$ & $    -99.00$ & $      0.89$ & $      0.75$ & $      3.38$ & $           9$ \\
$    -99.00$ & $    -99.00$ & $      0.38$ & $      1.73$ & $           4$ & $   -165.04$ & $           0$ & $      0.63$ & $      2.16$ & $           4$ & $    -64.05$ & $           0$ & $    -99.00$ & $      0.38$ & $      0.02$ & $      0.64$ & $           7$ \\
$    -99.00$ & $    -99.00$ & $      0.99$ & $      1.16$ & $           0$ & $     -2.48$ & $           0$ & $      0.14$ & $      1.23$ & $           1$ & $     -5.89$ & $           0$ & $    -99.00$ & $      1.27$ & $      0.49$ & $      0.99$ & $           7$ \\
$      4.38$ & $      0.86$ & $      6.62$ & $      1.81$ & $           4$ & $    -52.62$ & $           0$ & $      7.15$ & $      2.08$ & $           4$ & $    -18.62$ & $           0$ & $    -99.00$ & $      1.04$ & $      1.01$ & $      2.98$ & $           9$ \\
$    -99.00$ & $    -99.00$ & $     12.53$ & $      2.01$ & $           1$ & $     -2.08$ & $           0$ & $     11.24$ & $      2.38$ & $           4$ & $     -2.13$ & $           0$ & $    -99.00$ & $      0.72$ & $      0.60$ & $      0.62$ & $           7$ \\
$    -99.00$ & $    -99.00$ & $     -0.66$ & $      1.26$ & $           0$ & $     -4.26$ & $           0$ & $     -1.81$ & $      0.93$ & $           4$ & $     -3.01$ & $           0$ & $    -99.00$ & $      0.36$ & $      0.01$ & $      5.41$ & $           7$ \\
$    -99.00$ & $    -99.00$ & $      8.06$ & $      1.49$ & $           0$ & $     -2.60$ & $           0$ & $      5.71$ & $      2.05$ & $           1$ & $     -6.45$ & $           0$ & $    -99.00$ & $      2.07$ & $      2.61$ & $      1.09$ & $           7$ \\
$      2.15$ & $      0.58$ & $     -1.46$ & $      1.88$ & $           4$ & $    -20.43$ & $           0$ & $      0.97$ & $      1.92$ & $           4$ & $    -12.40$ & $           0$ & $    -99.00$ & $      0.60$ & $      0.27$ & $      3.14$ & $           8$ \\
$     57.82$ & $      3.56$ & $    160.17$ & $      7.57$ & $           0$ & $    -12.22$ & $           0$ & $    105.35$ & $      5.08$ & $           0$ & $     -6.01$ & $           0$ & $    -99.00$ & $      0.11$ & $      0.10$ & $    843.51$ & $           9$ \\
$      5.26$ & $      0.75$ & $      8.65$ & $      1.63$ & $           4$ & $    -23.95$ & $           0$ & $      5.08$ & $      1.35$ & $           0$ & $    -21.92$ & $           1$ & $    -99.00$ & $      0.86$ & $      0.71$ & $      3.80$ & $           8$ \\
$    -99.00$ & $    -99.00$ & $     -0.25$ & $      0.18$ & $           0$ & $     -9.92$ & $           0$ & $     -2.44$ & $      0.36$ & $           0$ & $     -5.65$ & $           2$ & $    -99.00$ & $      0.81$ & $      0.23$ & $      7.90$ & $           6$ \\
$    -99.00$ & $    -99.00$ & $      1.19$ & $      1.69$ & $           0$ & $     -2.47$ & $           0$ & $      1.16$ & $      1.96$ & $           1$ & $     -6.60$ & $           0$ & $    -99.00$ & $      1.50$ & $      2.20$ & $      2.10$ & $           7$ \\
$     31.55$ & $      1.40$ & $     42.23$ & $      2.67$ & $           0$ & $    -11.98$ & $           0$ & $     30.66$ & $      2.06$ & $           0$ & $     -6.04$ & $           0$ & $    -99.00$ & $      0.74$ & $      0.78$ & $      5.03$ & $           9$ \\
$     14.44$ & $      2.00$ & $     16.38$ & $      1.90$ & $           1$ & $     -2.63$ & $           0$ & $     11.74$ & $      1.80$ & $           1$ & $     -2.24$ & $           0$ & $    -99.00$ & $      0.99$ & $      0.97$ & $      3.23$ & $           8$ \\
$    -99.00$ & $    -99.00$ & $      0.88$ & $      1.70$ & $           4$ & $    -13.09$ & $           0$ & $     -0.15$ & $      0.01$ & $           4$ & $     -8.50$ & $           2$ & $    -99.00$ & $      0.49$ & $      0.40$ & $      1.07$ & $           6$ \\
$    -99.00$ & $    -99.00$ & $     -7.74$ & $      2.49$ & $           4$ & $    -15.10$ & $           2$ & $     -6.37$ & $      1.97$ & $           4$ & $    -15.97$ & $           2$ & $    -99.00$ & $      3.40$ & $      0.42$ & $      0.25$ & $           5$ \\
$    -99.00$ & $    -99.00$ & $     11.20$ & $      1.44$ & $           0$ & $     -2.48$ & $           0$ & $     15.69$ & $      1.93$ & $           0$ & $     -1.97$ & $           0$ & $    -99.00$ & $      2.41$ & $      1.71$ & $      0.34$ & $           7$ \\
$      7.41$ & $      1.66$ & $      6.22$ & $      1.43$ & $           0$ & $     -7.83$ & $           0$ & $      6.79$ & $      2.23$ & $           4$ & $     -9.51$ & $           0$ & $    -99.00$ & $      0.41$ & $      0.40$ & $      5.66$ & $           8$ \\
$      1.91$ & $      0.52$ & $      4.18$ & $      1.60$ & $           0$ & $     -2.05$ & $           0$ & $      1.47$ & $      2.08$ & $           1$ & $     -2.21$ & $           0$ & $    -99.00$ & $      0.68$ & $      0.69$ & $     17.49$ & $           9$ \\
$    -99.00$ & $    -99.00$ & $      1.17$ & $      1.31$ & $           0$ & $     -2.36$ & $           0$ & $      0.88$ & $      1.35$ & $           0$ & $     -1.77$ & $           0$ & $    -99.00$ & $      1.56$ & $      2.47$ & $      1.85$ & $           7$ \\
$      3.24$ & $      0.77$ & $      7.42$ & $      1.78$ & $           0$ & $     -2.60$ & $           0$ & $      4.82$ & $      1.67$ & $           0$ & $     -2.56$ & $           0$ & $    -99.00$ & $      0.68$ & $      0.75$ & $     17.36$ & $           9$ \\
$    -99.00$ & $    -99.00$ & $     -1.00$ & $      1.52$ & $           4$ & $     -7.91$ & $           0$ & $     -0.45$ & $      1.84$ & $           4$ & $     -3.44$ & $           0$ & $    -99.00$ & $      3.36$ & $      4.27$ & $      3.28$ & $           7$ \\
$    -99.00$ & $    -99.00$ & $      1.28$ & $      1.61$ & $           4$ & $    -28.68$ & $           0$ & $     -0.05$ & $      0.01$ & $           4$ & $    -71.67$ & $           2$ & $    -99.00$ & $      2.89$ & $      4.11$ & $      0.71$ & $           6$ \\
$    -99.00$ & $    -99.00$ & $      0.52$ & $      1.50$ & $           0$ & $     -2.37$ & $           0$ & $      1.20$ & $      1.53$ & $           0$ & $     -1.76$ & $           0$ & $    -99.00$ & $      1.65$ & $      0.26$ & $      0.58$ & $           7$ \\
$    -99.00$ & $    -99.00$ & $      4.05$ & $      0.19$ & $           0$ & $     -4.35$ & $           0$ & $     -0.81$ & $      2.11$ & $           4$ & $     -2.79$ & $           0$ & $    -99.00$ & $      0.80$ & $      0.69$ & $      2.62$ & $           7$ \\
$    -99.00$ & $    -99.00$ & $      2.48$ & $      0.46$ & $           1$ & $     -3.00$ & $           0$ & $      2.40$ & $      1.74$ & $           1$ & $     -2.15$ & $           0$ & $    -99.00$ & $      1.01$ & $      0.93$ & $      0.07$ & $           7$ \\
$    -99.00$ & $    -99.00$ & $     -0.23$ & $      1.43$ & $           1$ & $     -5.42$ & $           0$ & $     -0.88$ & $      1.96$ & $           1$ & $     -3.41$ & $           0$ & $    -99.00$ & $      1.43$ & $      0.07$ & $      0.94$ & $           7$ \\
$    -99.00$ & $    -99.00$ & $      5.58$ & $      1.68$ & $           0$ & $     -5.54$ & $           0$ & $      4.70$ & $      1.53$ & $           0$ & $     -3.29$ & $           0$ & $    -99.00$ & $      1.64$ & $      1.52$ & $      0.51$ & $           7$ \\
$     38.58$ & $      1.42$ & $     72.10$ & $      4.09$ & $           1$ & $     -4.56$ & $           0$ & $     63.25$ & $      3.88$ & $           1$ & $     -3.37$ & $           0$ & $    -99.00$ & $      0.99$ & $      1.03$ & $      5.93$ & $           9$ \\
$     37.73$ & $      3.57$ & $     13.42$ & $      2.10$ & $           1$ & $     -2.41$ & $           0$ & $      8.60$ & $      1.57$ & $           0$ & $     -2.36$ & $           0$ & $    -99.00$ & $      0.42$ & $      0.40$ & $     57.04$ & $           9$ \\
$      7.63$ & $      0.98$ & $      6.50$ & $      1.87$ & $           0$ & $   -211.86$ & $           0$ & $      3.16$ & $      1.54$ & $           0$ & $    -46.32$ & $           0$ & $    -99.00$ & $      0.31$ & $      0.22$ & $     44.47$ & $           9$ \\
$    -99.00$ & $    -99.00$ & $     -0.30$ & $      0.74$ & $           0$ & $     -4.93$ & $           0$ & $     -0.79$ & $      1.09$ & $           1$ & $     -3.33$ & $           0$ & $    -99.00$ & $      2.09$ & $      0.03$ & $      3.84$ & $           7$ \\
$    -99.00$ & $    -99.00$ & $      0.73$ & $      0.56$ & $           4$ & $    -18.94$ & $           0$ & $     -0.16$ & $      0.41$ & $           1$ & $     -9.67$ & $           0$ & $    -99.00$ & $      1.61$ & $      0.07$ & $      2.18$ & $           7$ \\
$    -99.00$ & $    -99.00$ & $      5.18$ & $      0.98$ & $           0$ & $     -2.37$ & $           0$ & $      3.98$ & $      1.42$ & $           0$ & $     -2.68$ & $           0$ & $    -99.00$ & $      0.58$ & $      0.13$ & $      0.69$ & $           7$ \\
$    -99.00$ & $    -99.00$ & $      4.63$ & $      1.83$ & $           4$ & $  -4159.75$ & $           0$ & $      3.92$ & $      1.44$ & $           0$ & $  -2565.74$ & $           0$ & $    -99.00$ & $      2.04$ & $      2.95$ & $      0.65$ & $           7$ \\
$     84.25$ & $      3.82$ & $     44.70$ & $      3.32$ & $           1$ & $     -2.00$ & $           0$ & $     35.51$ & $      3.30$ & $           1$ & $     -2.09$ & $           0$ & $    -99.00$ & $      0.24$ & $      0.23$ & $      3.94$ & $           9$ \\
$    -99.00$ & $    -99.00$ & $      0.38$ & $      1.58$ & $           1$ & $     -2.24$ & $           0$ & $      1.40$ & $      1.65$ & $           1$ & $     -2.27$ & $           0$ & $    -99.00$ & $      0.68$ & $      0.76$ & $      0.58$ & $           7$ \\
$     17.12$ & $      0.95$ & $     55.59$ & $      3.24$ & $           1$ & $    -14.21$ & $           0$ & $     36.57$ & $      2.75$ & $           1$ & $     -7.13$ & $           0$ & $    -99.00$ & $      0.05$ & $      4.16$ & $    146.62$ & $           9$ \\
$    -99.00$ & $    -99.00$ & $     -0.21$ & $      0.01$ & $           0$ & $     -1.81$ & $           2$ & $     -0.49$ & $      0.02$ & $           0$ & $     -1.77$ & $           2$ & $    -99.00$ & $      1.70$ & $      0.42$ & $      0.24$ & $           5$ \\
$     12.02$ & $      1.27$ & $     14.08$ & $      2.14$ & $           1$ & $     -2.26$ & $           0$ & $     15.92$ & $      2.35$ & $           4$ & $     -2.07$ & $           0$ & $    -99.00$ & $      0.96$ & $      1.07$ & $      4.64$ & $           9$ \\
$     84.16$ & $      0.96$ & $     51.44$ & $      2.83$ & $           0$ & $     -2.67$ & $           0$ & $     57.48$ & $      3.07$ & $           0$ & $     -2.32$ & $           0$ & $    -99.00$ & $      0.28$ & $      0.29$ & $      6.85$ & $           9$ \\
$      5.48$ & $      0.77$ & $      7.06$ & $      1.50$ & $           0$ & $     -2.44$ & $           0$ & $      6.06$ & $      1.33$ & $           0$ & $     -2.22$ & $           0$ & $    -99.00$ & $      0.66$ & $      0.68$ & $     10.13$ & $           9$ \\
$    -99.00$ & $    -99.00$ & $      1.56$ & $      0.09$ & $           4$ & $     -1.81$ & $           0$ & $     -0.77$ & $      1.04$ & $           4$ & $     -1.80$ & $           0$ & $    -99.00$ & $      1.91$ & $      0.07$ & $      3.32$ & $           7$ \\
$      4.00$ & $      0.77$ & $      0.53$ & $      1.51$ & $           4$ & $     -7.20$ & $           0$ & $      1.02$ & $      0.05$ & $           4$ & $     -8.10$ & $           0$ & $    -99.00$ & $      0.81$ & $      0.32$ & $     17.95$ & $           8$ \\
\enddata  
\tablecomments{Table 6 is published in its entirety in the machine-readable format. A portion is shown here for guidance regarding its form and content. (1) Unique object ID number, (2) object R.A. (J2000) in decimal degrees, (3) object decl. (J2000) in decimal degrees, (4) semimajor axis in the detection image, (5) ellipticity measured in the detection image, defined as $e=1-b/a$, where $b$  and $a$ are the semiminor and semimajor axes, respectively, (6) position angle measured in the detection image (degrees E from N), (7) Sextractor Kron radius used in AUTO photometry, (8) Sextractor half-light radius (FLUX\_RADIUS), (9,15,21,27,33) DECam $ugriz$ PSF fluxes appropriate for unresolved sources, (10,16,22,28,34) Errors on DECam $ugriz$ PSF fluxes, (11,17,23,29,35) DECam $ugriz$ total (Kron) fluxes, (12,18,24,30,36) Errors on DECam $ugriz$ total fluxes, (13,19,25,31,37) DECam $ugriz$ Sextractor internal flags, (14,20,26,32,38) DECam $ugriz$ Sextractor external flags, (39) $JK$ object ID number from \citet{geach17}, (40,42) $JK$ total (Kron) fluxes, (41,43) Errors on $JK$ total fluxes, (44,49) 3.6 and 4.5 $\mu$m IRAC Tractor total fluxes, (45,50)  3.6 and 4.5 $\mu$m IRAC Tractor errors, (46,51)  3.6 and 4.5 $\mu$m IRAC Tractor model profiles, (47,52) 3.6 and 4.5 $\mu$m IRAC Tractor log-probabilities, (48,49) 3.6 and 4.5 $\mu$m IRAC Tractor flags, (54) SDSS spectroscopic redshift, (55) EAZY peak photometric redshift, (56) EAZY $\chi^{2}$ minimum photometric redshift, (57)  EAZY $\chi^{2}$ minimum value, (58) EAZY number of filters used in photometric redshift estimate.}
\label{mcat4}
\end{deluxetable}
\clearpage
\end{landscape}

\end{document}